\documentclass{article}

\usepackage[english]{babel}

\usepackage[letterpaper,top=2cm,bottom=2cm,left=3cm,right=3cm,marginparwidth=1.75cm]{geometry}

\usepackage{natbib}
\usepackage{amsmath}
\usepackage{graphicx}
\usepackage[colorlinks=true, allcolors=blue]{hyperref}
\usepackage{multirow}
\usepackage{amssymb}
\usepackage{float}
\usepackage{longtable}
\usepackage[
singlelinecheck=false 
]{caption}
\usepackage{enumitem}
\usepackage{diagbox}
\usepackage{setspace}
\usepackage{slashbox}
\usepackage{booktabs,subcaption,amsfonts,dcolumn}
\usepackage{tablefootnote}
\newcolumntype{d}[1]{D..{#1}}

\usepackage[markup=underlined,dvipsnames,svgnames,x11names]{xcolor}
\usepackage{changes}

\title{Exact Methods of Homogeneity Test of Proportions for Bilateral and Unilateral Correlated Data}
\author{Shuyi Liang and Chang-Xing Ma \thanks{\href{mailto:cxma@buffalo.edu}{cxma@buffalo.edu}}}
\definechangesauthor[color=BrickRed]{AU}
\usepackage{todonotes}
\setcommentmarkup{\todo[color={authorcolor!20},size=\scriptsize]{#3: #1}}

\providecommand{\keywords}[1]
{
  \small	
  \textbf{\text{Keywords:}} #1
}

\begin{document}

\maketitle
Department of Biostatistics, University at Buffalo, Buffalo, New York 14214, USA

\begin{abstract}

Subjects in clinical studies that investigate paired body parts can carry a disease on either both sides (bilateral) or a single side (unilateral) of the organs. Data in such studies may consist of both bilateral and unilateral records. However, the correlation between the paired organs is often ignored, which may lead to biased interpretations. Recent literatures have taken the correlation into account. For example, Ma and Wang \citeyearpar{ma2021testing} proposed three asymptotic procedures for testing the homogeneity of proportions of multiple groups using combined bilateral and unilateral data and recommended the score test. It is of importance to notice that the asymptotic behavior is not guaranteed if the sample size is small, resulting in uncontrolled type I error rates. In this paper, we extend their work by considering exact approaches and compare these methods with the score test proposed by Ma and Wang \citeyearpar{ma2021testing} in terms of type I errors and statistical powers. Additionally, two real-world examples are used to illustrate the application of the proposed approaches.

\end{abstract}

\keywords{constant R model; homogeneity test; nuisance parameter; unilateral and bilateral data; exact method}

\section{Introduction}

Clinical trials that involve collecting binary measurements, such as the presence or absence of a disease, from paired parts of a human body (e.g., ears and eyes) require comprehensive data collections and careful statistical analyses. Because of the latent correlation between the paired organs, ignoring this hidden characteristic will lead to biased inferences (Rosner \citeyear{rosner1982statistical}, Dallal \citeyear{dallal1988paired}, and Donner and Banting \citeyear{donner1988analysis}). Zhang and Ying (\citeyear{zhang2018statistical}) evaluated statistical practices in the analyses of eye data published in the British Journal of Ophthalmology (BJO) in 1995 and 2017. Their findings revealed that a majority of the studies did not account for the intereye correlation and there had been no significant change in data collection methods over decades. 

Taking the intercorrelation into account, Rosner \citeyearpar{rosner1982statistical} introduced a so-called "constant R" model, assuming that the conditional probability of observing a disease at one organ given a disease at the other site is constantly proportional to the prevalence of the disease. For studies in which data are available at both sites of the paired body parts, Tang et al. \citeyearpar{tang2008testing} presented eight statistical tests to evaluate the equality of prevalence between two groups. Ma et al. \citeyearpar{ma2015homogeneity} extended their work by considering the scenarios with more than two groups and proposed three asymptotic test statistics for testing the homogeneity of prevalence given bilateral data. Although asymptotic methods generally provide satisfactory controls over type I errors when the sample size is large enough and the normal approximation is guaranteed, the asymptotic approximations work poorly where the number of subjects is insufficient (Agresti \citeyear{agresti1992survey}; Storer and Kim \citeyear{storer1990exact}). To overcome the limitations associated with asymptotic procedures, it is advisable to consider exact methods that offer more precise and accurate results. Under exact frameworks, it is straightforward to determine the exact p-values if the probability distribution is known, such as in the case of the hypergeometric distribution where both margins are fixed in a contingency table. However, the calculations of p-values become challenging when there are unknown nuisance parameters involved, even if the form of distribution is known. There are numerous methods that have been developed to eliminate nuisance parameters. Storer and Kim \citeyearpar{storer1990exact} considered replacing the nuisance
parameters with their maximum likelihood estimations (MLEs) under the null hypothesis, which is referred to as the E approach. Basu \citeyearpar{Basu1977OnTE} proposed the M approach to calculate the p-value by maximizing the probability of observing the tail area over the whole parameter space. In contrast to considering the entire parameter space, which often results in conservative p-values close to 1, Berger and Boos \citeyearpar{berger1994p} maximized the probability of the tail area over a confidence interval of the nuisance parameter. Lloyd \citeyearpar{lloyd2008exact} utilized the p-value from the E approach to find the tail area and then maximized the probability of the area within the whole nuisance parameter space, which is referred to as the E+M approach. Tang et al. \citeyearpar{tang2006statistical} employed the E and M approaches for testing the equivalence of prevalence of two groups. In addition to the previous two methods, Shan and Ma \citeyearpar{shan2014exact} further investigated the E+M approach and a conditional method assuming both margins of a contingency table are fixed. 
Liu et al. \citeyearpar{liu2017exact} considered exact tests of homogeneity of prevalence when there are more than two groups.

It is of importance to note that not all of the participants have the capability to provide measurements for both organs. For example, individuals with acute otitis media with effusion may carry the disease in either both ears or only a single ear at the beginning of the study. Therefore, data collected in these studies consist of a combination of both bilateral and unilateral data. With combined data, Ma and Wang \citeyearpar{ma2021testing} examined asymptotic approaches for testing the equality of multiple proportions using the "constant R" model. Qiu et al. \citeyearpar{qiu2022confidence} developed several confidence intervals to assess the equivalence of two treatments with asymptotic and bootstrap resampling methods. Similar to the studies that solely focus on bilateral data, the performance of asymptotic procedures heavily relies on large sample sizes. However, the exploration of exact procedures specifically designed for combined studies remains an open area for further research and investigation.

In this article, we introduce five exact approaches as alternative methods to the asymptotic procedures proposed by Ma and Wang \citeyearpar{ma2021testing} for conducting homogeneity tests of prevalence among multiple groups. In Section 2, we provide a brief introduction to the constant R model and outline the existing asymptotic method. Section 3 delves into the details of the five proposed exact approaches. In Section 4, we present numerical studies comparing these methods in terms of their controls of type I errors and statistical powers. To provide practical insight, we apply all approaches, including the existing asymptotic approach, to two real examples in Section 5. Finally, we summarize our conclusions in Section 6.

\section{Data Structure and Existing Method}
\subsection{Constant R model}
Assume $M$ subjects contributing bilateral data and $N$ subjects contributing unilateral data are randomized to $g$ groups. In the unilateral cohort, let $n_{ir}$ represent the number of subjects with $r$ response(s) in the $i$th group, where $i=1, 2, ..., g$ and $r=0, 1$. Similarly, let $m_{ir^*}$ denote the number of subjects with $r^*$ response(s) in the $i$th group in the bilateral cohort, where $i=1, 2, ..., g$ and $r^*=0, 1, 2$. Group totals are defined as follows: \(n_{i}=\sum_{r=0}^{1} n_{ir}\) for the unilateral cohort and \(m_{i}=\sum_{r^*=0}^{2} m_{ir^*}\) for the bilateral cohort. Furthermore, the total number of subjects with $r$ response(s) are denoted as $N_r$ in the unilateral cohort and $S_{r^*}$ in the bilateral cohort. The data structure is presented in \hyperref[tab:datastructure]{Table \ref*{tab:datastructure}}. It is reasonable to assume that $(n_{i0}, n_{i1})^{T}$ follows a binomial distribution $Bin(n_{i}; P_{ui0}, P_{ui1})$, where $P_{uir}$ corresponds to the probability of a subject from the $ith$ group having $r$ response(s) $(r=0, 1)$ in the unilateral cohort. For the bilateral cohort, we assume $(m_{i0}, m_{i1}, m_{i2})^{T}$ follows a multinomial distribution $Multi(m_{i}; P_{bi0}, P_{bi1}, P_{bi2})$.\\
\begin{table}[h]
\centering
\caption{Data Layout}
\label{tab:datastructure}
\begin{tabular}{l|cccc||ccc}
\toprule
\multirow{3}{*}{group ($i$)} & \multicolumn{4}{c||}{Bilateral} & \multicolumn{3}{c}{Unilateral}\\ \cline{2-8} 
& \multicolumn{3}{c|}{Response ($r^*$)}& \multirow{2}{*}{Total} & \multicolumn{2}{c|}{Response ($r$)}& \multirow{2}{*}{Total} \\ \cline{2-4} \cline{6-7}
& \multicolumn{1}{c}{0}         & \multicolumn{1}{c}{1}         & \multicolumn{1}{c|}{2}         &                        & \multicolumn{1}{c}{0}         & \multicolumn{1}{c|}{1}         &                        \\ \hline
1& \multicolumn{1}{c}{$m_{10}$} & \multicolumn{1}{c}{$m_{11}$} & \multicolumn{1}{c|}{$m_{12}$} & $m_{1}$               & \multicolumn{1}{c}{$n_{10}$} & \multicolumn{1}{c|}{$n_{11}$} & $n_{1}$               \\ 
2& \multicolumn{1}{c}{$m_{20}$} & \multicolumn{1}{c}{$m_{21}$} & \multicolumn{1}{c|}{$m_{22}$} & $m_{2}$               & \multicolumn{1}{c}{$n_{20}$} & \multicolumn{1}{c|}{$n_{21}$} & $n_{2}$               \\ 
...  & \multicolumn{3}{c|}{...} & ... & \multicolumn{2}{c|}{...} & ... \\ 
g& \multicolumn{1}{c}{$m_{g0}$} & \multicolumn{1}{c}{$m_{g1}$} & \multicolumn{1}{c|}{$m_{g2}$} & $m_{g}$               & \multicolumn{1}{c}{$n_{g0}$} & \multicolumn{1}{c|}{$n_{g1}$} & $n_{g}$               \\ \hline
Total& \multicolumn{1}{c}{$S_{0}$} & \multicolumn{1}{c}{$S_{1}$} & \multicolumn{1}{c|}{$S_{2}$} & $M$               & \multicolumn{1}{c}{$N_{0}$} & \multicolumn{1}{c|}{$N_{1}$} & $N$               \\   
\bottomrule
\end{tabular}

\end{table}

Let the probability of observing a response on the $k$th site of the $j$th subject from the $i$th group be $\pi_{i}$. Define $Z_{ijk}=1 (k=1, 2)$ as the occurrence of such an event. Under the assumption by Rosner \citeyearpar{rosner1982statistical}, the conditional probability of observing a response at one site given a response at the other site of a subject is proportional to $\pi_i$ by a constant $R$, That is, 
$Pr(Z_{ijk}=1)=\pi_{i}$ and $Pr(Z_{ijk}=1|Z_{ij(3-k)}=1)=\pi_{i}R \;$,
where $0< \pi_{i}<1$. Therefore, $P_{ui0}, P_{ui1}, P_{bi0}, P_{bi1},$ and $P_{bi2}$ can be expressed as functions of $\pi_i$ and $R$:
\[\begin{cases}
     \text{$P_{ui0}=1-\pi_i$}\\
     \text{$P_{ui1}=\pi_i$}\\
     \text{$P_{bi0}=1-2\pi_i+R\pi_i^2$}\\
     \text{$P_{bi1}=2(\pi_i-R\pi_i^2)$}\\
     \text{$P_{bi2}=R\pi_i^2$.}\\
    \end{cases} \]\\
The focus lies on testing the equality of response rates among the $g$ groups. As a result, the null and the alternative hypotheses can be formulated as follows:
\[\begin{cases}
     \text{$H_0$: $\pi_1=\pi_2=...=\pi_g=\pi$}\\
    \text{$H_a$: $\pi_i\neq \pi_j$ for some $i\neq j$, where $i,j \in \{1, 2, ..., g\}$},
    \end{cases} \]
where $\pi$ is an unknown quantity.

\subsection{Existing Method}
Denote the observed data as $M^*=(m_{10},m_{11},m_{12},n_{10},n_{11},...,m_{g0},m_{g1},m_{g2},n_{g0},n_{g1})$. Under the assumption of independence between different groups or different cohorts, the likelihood function can be expressed as
\begin{flalign*}
\mathcal{L}(M^*)  =\prod_{i=1}^{g}\frac{m_{i}!}{m_{i0}!m_{i1}!m_{i2}!}P_{bi0}^{m_{i0}}P_{bi1}^{m_{i1}}P_{bi2}^{m_{i2}}\times \frac{n_{i}!}{n_{i0}!n_{i1}!}P_{ui0}^{n_{i0}}P_{ui1}^{n_{i1}}.
\end{flalign*}
And the corresponding log-likelihood function in terms of $\pi_1, \pi_2, ..., \pi_g,$ and $R$ is given by
\begin{flalign*}
l(\pi_1, \pi_2, ..., \pi_g, R) = \sum_{i=1}^{g} & \left[ m_{i0} \log(R\pi_i^2 - 2\pi_i + 1) + m_{i1} \log(2\pi_i - 2R\pi_i^2) + m_{i2} \log(R\pi_i^2) \right. \\
& \left. + n_{i0} \log(1 - \pi_i) + n_{i1} \log(\pi_i) \right].
\end{flalign*}

Under the null hypothesis, the parameters $R$ and $\pi$ satisfy the condition $max\{0, (2-1/\pi)/\pi\}<R<1/\pi$ and the log-likelihood function can be further simplified as 

\begin{flalign*}
l(\pi, R) = &  S_{0} \log(R\pi^2 - 2\pi + 1) + S_{1} \log(2\pi - 2R\pi^2) + S_{2} \log(R\pi^2)  \\
&  + N_{0} \log(1 - \pi) + N_{1} \log(\pi) .
\end{flalign*}

\subsubsection{Maximum Likelihood Estimates (MLEs) under $H_0$}
Constrained Maximum Likelihood Estimates (MLEs) in the parameter space under $H_0$ can be obtained by solving the following two equations
\[\begin{cases}
     \text{$\frac{\partial l}{\partial \pi}=\frac{N_1 }{\pi }+\frac{2\,S_2 }{\pi }+\frac{N_0 }{\pi -1}+\frac{S_0 \,{\left(2\,R\,\pi -2\right)}}{R\,{\pi }^2 -2\,\pi +1}-\frac{S_1 \,{\left(4\,R\,\pi -2\right)}}{2\,\pi -2\,R\,{\pi }^2 }=0$}\\
    \text{$\frac{\partial l}{\partial R}=\frac{S_2 }{R}+\frac{S_0 \,{\pi }^2 }{R\,{\pi }^2 -2\,\pi +1}-\frac{S_1 \,\pi }{1 -R\,\pi }=0$.}
    \end{cases} \]\\
Ma and Wang \citeyearpar{ma2021testing} derived the closed-form solutions for the contrained MLEs, denoted as $\hat{\pi}$ and $\hat{R}$. For more comprehensive information, we recommend referring to their work.

\subsubsection{Score Test ($T_{SC}$)}
Define $U=(U_1,U_2,...,U_g,U_{g+1})=(\frac{\partial{l}}{\partial{\pi_{1}}}, \frac{\partial{l}}{\partial{\pi_{2}}},...,\frac{\partial{l}}{\partial{\pi_{g}}},\frac{\partial{l}}{\partial{R}})$   Then the score test statistic $T_{SC}$ can be expressed as 
\begin{flalign*}
T_{SC}(M^*)=U\mathcal{I}(\pi_1,\pi_2,...,\pi_g,R)^{-1}U^T|\pi_1=\pi_2=...=\pi_g=\hat{\pi},R=\hat{R}
\end{flalign*}
where $\mathcal{I}(\pi_1,\pi_2,...,\pi_g,R)$ is the Fisher information under the alternative hypothesis. Ma and Wang \citeyearpar{ma2021testing} simplified the form of $T_{SC}$ as the following:
\begin{flalign*}
T_{SC}(M^*)=\sum_{i=1}^{g}\frac{U_i^2}{I_{ii}}+\left( \sum_{i=1}^{g}\frac{I_{i,g+1}U_i}{I_{ii}} \right)^2 \left( I_{g+1,g+1}-\sum_{k=1}^{g}\frac{I_{k,g+1}^2}{I_{kk}} \right)^{-1},
\end{flalign*}
where, $I_{ii}, I_{i,g+1},$ and $I_{g+1,g+1}$ are the elements of the Fisher information. Again, we kindly suggest that readers refer to Ma and Wang \citeyearpar{ma2021testing} for detailed formulas of $\mathcal{I}(\pi_1,\pi_2,...,\pi_g,R)$. Under the null hypothesis, the score test statistic follows asymptotically a chi-sqaure distribution with $g-1$ degrees of freedom according to Rao \citeyearpar{rao1948large}. The asymptotic p-value can be calculated as 
\begin{flalign*}
P_A(M^*)=Prob(X>T_{SC}(M^*)),
\end{flalign*}
where $X$ follows $\mathcal{X}^2_{g-1}$. The asymptotic behavior of the score test statistic is valid when the sample size is sufficiently large. However, the approximation performs inadequately and cannot control type I errors when the sample size is small. This deficiency is demonstrated in Section 4. To address this concern, we propose five exact methods as alternatives to overcome the limitations associated with the asymptotic approach. Four of the approaches in this study are based on the score test statistic. The choice of the score test statistic was made due to its explicit form, which helps alleviate the computational complexity.

\section{Exact Methods}
Exact p-values can be computed by utilizing the probability models $(n_{i0}, n_{i1})^{T} \sim Bin(n_{i}; P_{ui0}, P_{ui1})$ and $(m_{i0}, m_{i1}, m_{i2})^{T} \sim Multi(m_{i}; P_{bi0}, P_{bi1}, P_{bi2})$ given fixed $(m_1,n_1,m_2,n_2,...,m_g,n_g)$. However, the presence of two nuisance parameters, $\pi$ and $R$, poses a challenge to these computations since the true parameter values are unkown. To address this difficulty, the following approaches are proposed.
\subsection{E Approach}
The E approach simply replace $\pi$ and $R$ with their constrained MLEs $\hat{\pi}$ and $\hat{R}$, respectively. Hence, the p-value is defined by
\begin{flalign*}
P_E(M^*)=Prob(X\geq T_{SC}(M^*)|\pi=\hat{\pi},R=\hat{R})=\sum_{M \in \Omega_E(M^*)}^{}\mathcal{L}(M|\pi=\hat{\pi},R=\hat{R}),
\end{flalign*}
where $X$ follows $\mathcal{X}^2_{g-1}$ and $\Omega_E(M^*)=\{M|T_{SC}(M)\geq T_{SC}(M^*) \}$ represents the tail area of the observed data $M^*$. 

\subsection{M Approach}
The second approach, M approach, focuses on maximizing the above summation of probabilities over the entire parameter space. The exact p-value based on the M approach is give by
\begin{flalign*}
P_M(M^*)=\sup_{\pi \in (0,1), max\{0, \frac{2-\frac{1}{\pi}}{\pi}\}<R<\frac{1}{\pi}}  \Biggl\{ \sum_{M \in \Omega_M(M^*)}^{}\mathcal{L}(M;\pi,R) \Biggl\}.
\end{flalign*}
The tail area is defined as $\Omega_M(M^*)=\{M|T_{SC}(M)\geq T_{SC}(M^*) \}$, which is the same as the E approach.

\subsection{E+M Approach}
Rather than directly defining the tail area using the score test statistic, an alternative method, the E+M approach, considers finding the tail area based on the p-values calculated by the E approach and defining the p-value as 
\begin{flalign*}
P_{E+M}(M^*)=\sup_{\pi \in (0,1), max\{0, \frac{2-\frac{1}{\pi}}{\pi}\}<R<\frac{1}{\pi}}  \Biggl\{ \sum_{M \in \Omega_{E+M}(M^*)}^{}\mathcal{L}(M;\pi,R) \Biggl\},
\end{flalign*}
where $\Omega_{E+M}(M^*)=\{M|P_E(M)\leq P_E(M^*) \}$.

\subsection{CI Approach}
Next, we will discuss the CI approach, which is based on confidence intervals (CIs) according to Berger and Boos \citeyearpar{berger1994p} and Silvapulle \citeyearpar{silvapulle1996test}. In contrast to the M approach, which considers the entire parameter space and may result in an excessively large p-value due to the use of supremum, the CI approach takes into account parameter values that fall within the confidence intervals of $\pi$ and $R$ under the null hypothesis. The CI-type p-value is defined by
\begin{flalign*}
P_{CI}(M^*)=\sup_{\pi \in CI_{\pi}, R \in CI_R}  \Biggl\{ \sum_{M \in \Omega_M(M^*)}^{}\mathcal{L}(M;\pi,R) \Biggl\}+3\beta,
\end{flalign*}
where $CI_{\pi}$ and $CI_R$ are confidence intervals of $\pi$ and $R$, respectively. The confidence level of both $CI_{\pi}$ and $CI_R$ is denoted as $1-\beta$. A statistic "p-value" is called a valid p-value if the property $Prob(p-value \leq \alpha|H_0)\leq \alpha$ is preserved (Vexler \citeyear{vexler2021valid}). It is straightforward to show that the aforementioned $P_{CI}(M^*)$ is a valid p-value. The following is a brief proof:

\begin{flalign*}
Prob(P_{CI}(M^*)\leq \alpha|H_0)=&Prob(P_{CI}(M^*)\leq \alpha, \pi_0 \in CI_{\pi}, R_0 \in CI_R|H_0)\\
&+Prob(P_{CI}(M^*)\leq \alpha, \pi_0 \in CI_{\pi}, R_0 \notin CI_R|H_0)\\
&+Prob(P_{CI}(M^*)\leq \alpha, \pi_0 \notin CI_{\pi}, R_0 \in CI_R|H_0)\\
&+Prob(P_{CI}(M^*)\leq \alpha, \pi_0 \notin CI_{\pi}, R_0 \notin CI_R|H_0)\\
\leq&Prob(P_{CI}(M^*)\leq \alpha, \pi_0 \in CI_{\pi}, R_0 \in CI_R|H_0)+Prob(R_0 \notin CI_{R}|H_0)\\
&+Prob(\pi_0 \notin CI_{\pi}|H_0)+Prob(\pi_0 \notin CI_{\pi},R_0 \notin CI_R|H_0)\\
\leq&Prob(Prob(M\in \Omega_M(M^*)|\pi_0,R_0)+3\beta \leq \alpha, \pi_0 \in CI_{\pi}, R_0 \in CI_R|H_0)\\
&+Prob(R_0 \notin CI_{R}|H_0)+2\times Prob(\pi_0 \notin CI_{\pi}|H_0)\\
\leq&Prob(Prob(M\in \Omega_M(M^*)|\pi_0,R_0)\leq  \alpha-3\beta|H_0)+3\beta\\
=&\alpha-3\beta+3\beta\\
=&\alpha,
\end{flalign*}
where $Prob(M\in \Omega_M(M^*)|\pi_0,R_0)$ is uniformly distributed on $[0,1]$ under $H_0$ and $\pi_0$ and $R_0$ are the true but unknown parameters.

To construct confidence intervals for $\pi$ and $R$, we derive another score test statistic ($T_{SC}^*$) under the null hypothesis $H_0:\pi_1=\pi_2=...=\pi_g=\pi$. Given $R$ is known, the conditional MLE of $\pi$ can be obtained by solving the quartic equation derived by Ma and Wang \citeyearpar{ma2021testing}. Similarly, the conditional MLE of $R$ given $\pi$ is known, can be obtained by one of the roots of the quadratic equation according to Ma and Wang \citeyearpar{ma2021testing}. Let $U_C=(\frac{\partial{l}}{\partial{\pi}},\frac{\partial{l}}{\partial{R}})$ and $\mathcal{I}_C(\pi,R)$ denote the Fisher information matrix under $H_0$. Therefore, the new score test statistic can be expressed as:
\begin{flalign*}
T_{SC}^*(M^*)=U_C\mathcal{I}_C(\pi,R)^{-1}U_C^T,
\end{flalign*}
where
\[
\mathcal{I}_C(\pi,R)=
\begin{bmatrix}
I_{11} & I_{12}\\
I_{21} & I_{22}\\
\end{bmatrix} 
\]
and
\begin{flalign*}
I_{11}&=E(-\frac{\partial^2{l}}{\partial{\pi^2}})=\frac{N}{\pi }-\frac{N}{\pi -1}+4\,M\,R+\frac{4\,M\,{{\left(\pi \,R-1\right)}}^2 }{R\,\pi^2 -2\,\pi +1}-\frac{2\,M\,{{\left(2\,\pi \,R-1\right)}}^2 }{\pi \,{\left(\pi \,R-1\right)}},\\
I_{12}&=I_{21}=E(-\frac{\partial^2{l}}{\partial{\pi}\partial{R}})=-\frac{2\,M\,\pi^2 \,{\left(R-1\right)}}{{\left(\pi \,R-1\right)}\,{\left(R\,\pi^2 -2\,\pi +1\right)}},\\
I_{22}&=E(-\frac{\partial^2{l}}{\partial{R^2}})=-\frac{M\,\pi^2 \,{\left(\pi \,R-2\,\pi +1\right)}}{R\,{\left(\pi \,R-1\right)}\,{\left(R\,\pi^2 -2\,\pi +1\right)}}.\\
\end{flalign*}

The upper limit of the CI of $\pi$ can be found via the following procedures:\\
(1) Set the initial value of $\pi$ as $\hat{\pi}$, where $\hat{\pi}$ is the constrained MLE of $\pi$ under $H_0$. Let flag = 1 and stepsize = $min\{0.01,(1-\pi^{(t)})/10\}$;\\
(2) Update  $\pi^{(t+1)}=\pi^{(t)}$ + flag $\times$ stepsize and compute the conditional MLE of $R$, denoted as $\Tilde{R}$, given $\pi^{(t+1)}$ by solving the quadratic equation derived by Ma and Wang \citeyearpar{ma2021testing}. Compute the score test statistic $(T_{SC}^*(M^*))^{(t+1)}=U_C\mathcal{I}_C(\pi,R)^{-1}U_C^T|\pi=\pi^{(t+1)},R=\Tilde{R}$;\\
(3) If $(T_{SC}^*(M^*))^{(t+1)} > \mathcal{X}^2_{1, 1-\beta}$, update the search direction by letting flag = -1 and change stepsize = stepsize $\times 1/\pi$, then return to step (2). Otherwise, let flag = 1 and return to step (2);\\
(4) Repeat steps (2) and (3) until convergence, i.e., the stepsize is sufficiently small (e.g., $10^{-4}$).

The lower limit of the CI of $\pi$ can be found by initializing flag = -1 in step (1) and then update it to 1 if $(T_{SC}^*(M^*))^{(t+1)} > \mathcal{X}^2_{1, 1-\beta}$ or -1 if $(T_{SC}^*(M^*))^{(t+1)} \leq \mathcal{X}^2_{1, 1-\beta}$.

In a similar manner, the upper bound of the CI of $R$ can be obtained by the following procedures:\\
(1) Set the initial value of $R$ as $\hat{R}$, where $\hat{R}$ is the constrained MLE of $R$ under $H_0$. Let flag = 1 and stepsize = 0.1;\\
(2) Update  $R^{(t+1)}=R^{(t)}$ + flag $\times$ stepsize and compute the conditional MLE of $\pi$, denoted as $\Tilde{\pi}$, given $R^{(t+1)}$ by solving the quartic equation derived by Ma and Wang \citeyearpar{ma2021testing}. Compute the score test statistic $(T_{SC}^*(M^*))^{(t+1)}=U_C\mathcal{I}_C(\pi,R)^{-1}U_C^T|R=R^{(t+1)},\pi=\Tilde{\pi}$;\\
(3) If $(T_{SC}^*(M^*))^{(t+1)} > \mathcal{X}^2_{1, 1-\beta}$, update the search direction by letting flag = -1 and change stepsize = stepsize $\times 1/\pi$, then return to step (2). Otherwise, let flag = 1 and return to step (2);\\
(4) Repeat steps (2) and (3) until convergence, i.e., the stepsize is sufficiently small (e.g., $10^{-4}$).

The lower bound of the CI of $R$ can be found by initializing flag = -1 in step (1) and then update it to 1 if $(T_{SC}^*(M^*))^{(t+1)} > \mathcal{X}^2_{1, 1-\beta}$ or -1 if $(T_{SC}^*(M^*))^{(t+1)} \leq \mathcal{X}^2_{1, 1-\beta}$.

\subsection{C Approach}
The C approach assumes both margins $(m_1,n_1,m_2,n_2,...,m_g,n_g)$ and $(S_0,S_1,S_2,N_0,N_1)$ are fixed. Hence, all the possible observations are limited to 
\begin{flalign*}
\Phi(M^*)=\{M|S_0=S_0^*,S_1=S_1^*,S_2=S_2^*,N_0=N_0^*,N_1=N_1^*,m_i=m_i^*,n_i=n_i^*,i=1,2,...,g \}.
\end{flalign*}
The tail area is given by
\begin{flalign*}
\Omega_{C}(M^*)=\{M|M \in \Phi(M^*), T_{SC}(M)\geq T_{SC}(M^*) \}.
\end{flalign*}
The exact p-value is then computed with the following formula:
\begin{flalign*}
P_C(M^*)=\sum_{M \in \Omega_{C}(M^*)}^{}\Biggl(\frac{\prod_{i=1}^{g}\frac{m_i!}{m_{i0}!m_{i1}!m_{i2}!}}{\frac{M!}{S_0!S_1!S_2!}} \times \frac{\prod_{i=1}^{g}\frac{n_i!}{n_{i0}!n_{i1}!}}{\frac{N!}{N_0!N_1!}} \Biggl)
\end{flalign*}

\section{Numerical Study}
In this section, we evaluate the performances of the proposed exact methods discussed in previous sections with different sample sizes. Unlike simulation-based approaches, exact tests consider all possible scenarios. As a result, no simulation will be performed. In real-world analyses, particularly in studies involving rare diseases, the number of participants is often limited, regardless of the number of groups. Hence, type I errors and powers are investigated for cases where the total sample size $M+N$ is approximately 30 and 20. We focus on balanced designs with $m_1=m_2=...=m_g$ and $n_1=n_2=...=n_g$. Unbalanced studies can be investigated in a similar way and therefore are omitted in this article. 

\begin{figure}[H]
        \caption{Surface plots of type I errors for $g=2$, $m_i=10$, and $n_i=5$ (A and E approaches, $i=1,2$)}
        \label{fig:surf_g2_m10_n5_A_E}
     \centering

         \includegraphics[width=\textwidth]{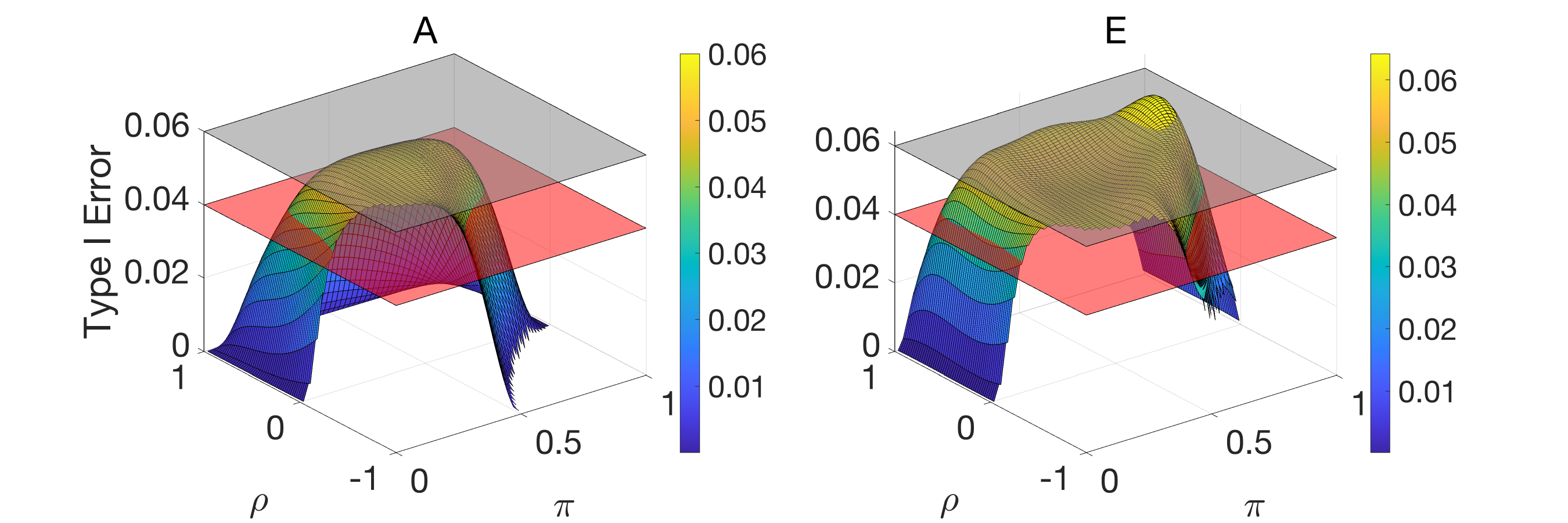}

\end{figure}

\begin{figure}[H]
        \caption{Surface plots of type I errors for $g=4$, $m_i=4$, and $n_i=3$ (C approach, $i=1,2,3,4$)}
        \label{fig:surf_g4_m4_n3_C}
     \centering

         \includegraphics[scale=0.35]{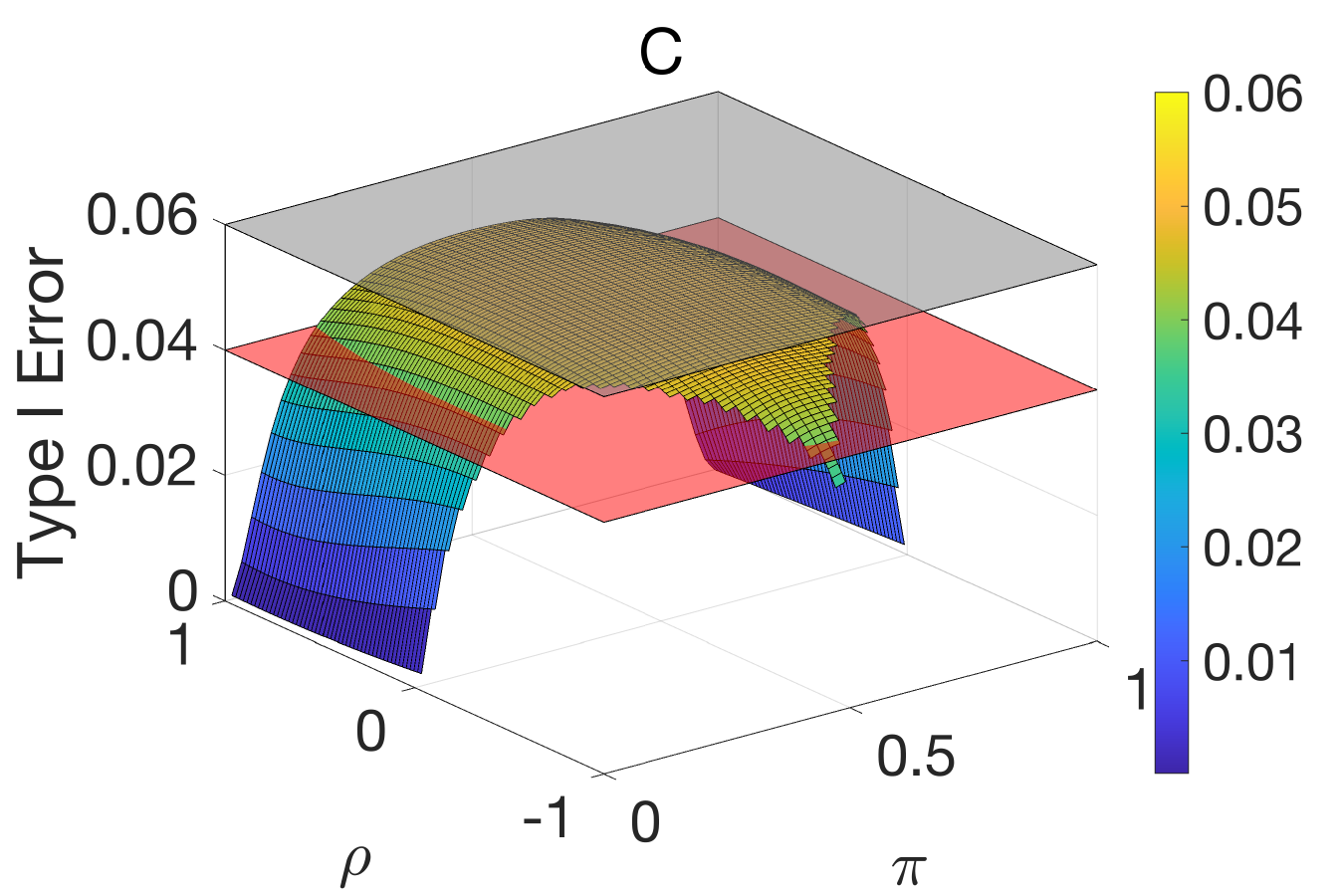}

\end{figure}

\begin{figure}[H]
        \caption{Surface plots of type I errors for $g=2$, $m_i=5$, and $n_i=5$ (A and E approaches, $i=1,2$)}
        \label{fig:surf_g2_m5_n5_A_E}
     \centering

         \includegraphics[width=\textwidth]{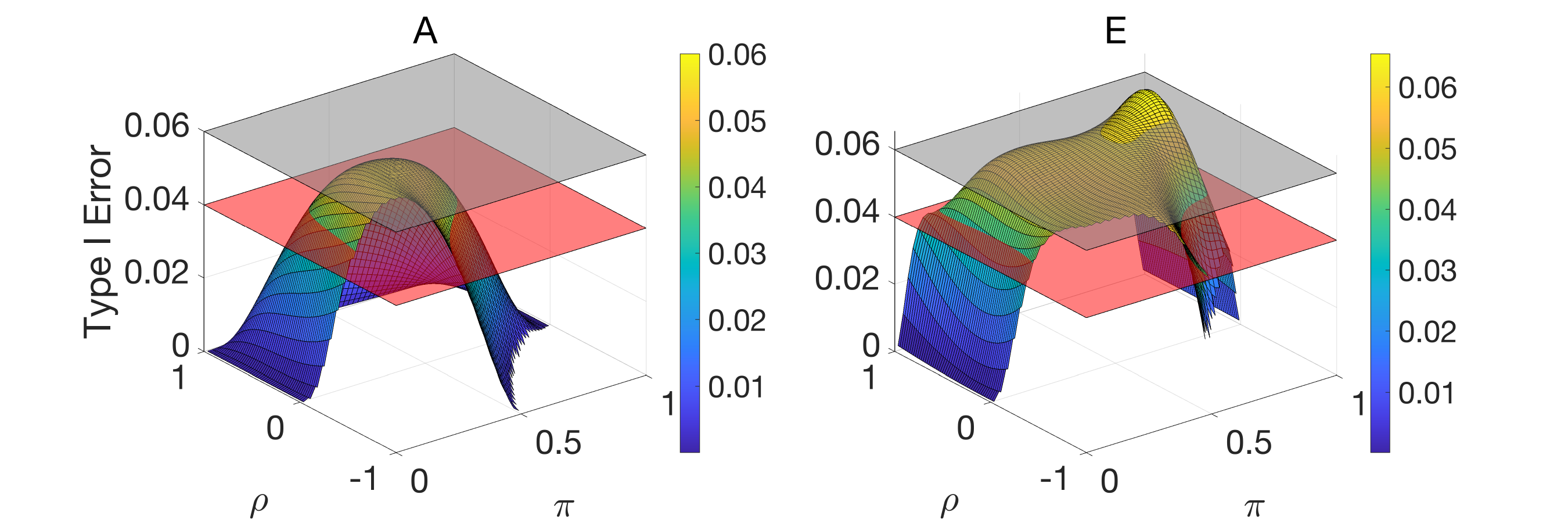}

\end{figure}

\begin{figure}[H]
        \caption{Surface plots of type I errors for $g=4$, $m_i=3$, and $n_i=2$ (C and CI approaches, $i=1,2,3,4$)}
        \label{fig:surf_g4_m3_n2_C_CI}
     \centering

         \includegraphics[width=\textwidth]{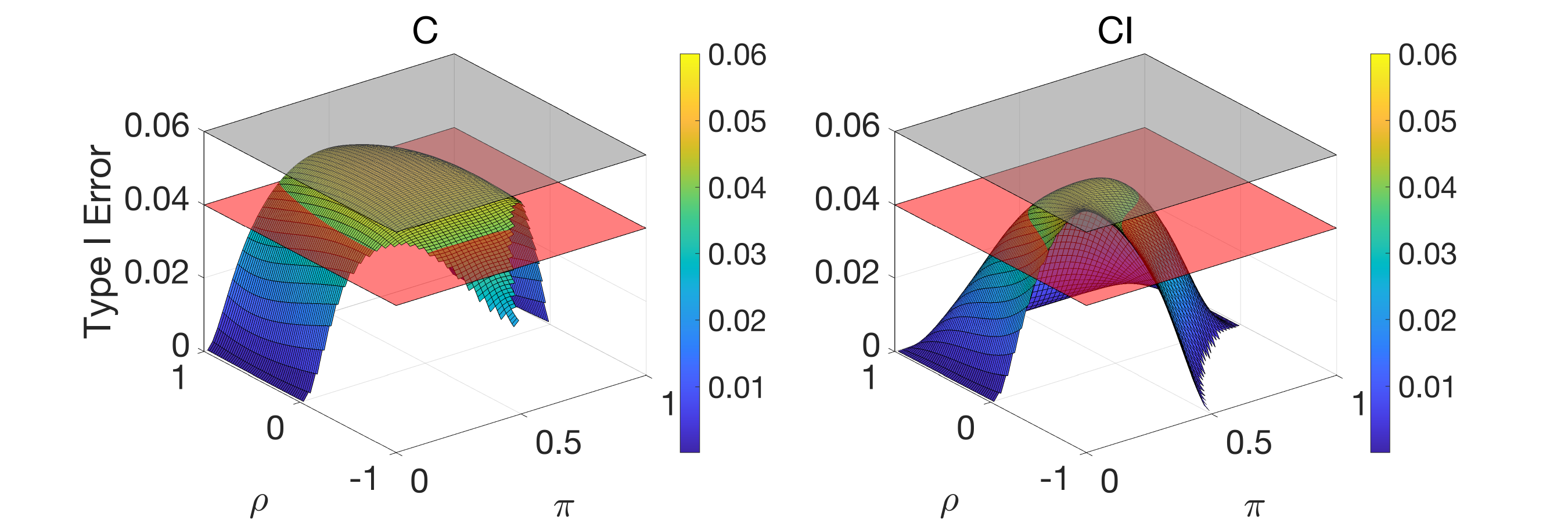}

\end{figure}

Given known $\pi$ and $\rho$, where 
\begin{flalign*}
\rho=\frac{cov(Z_{ijk}, Z_{ij(3-k})}{\sqrt{var(Z_{ijk})var(Z_{ij(3-k)})}}=\frac{\pi R -\pi}{1-\pi},
\end{flalign*}
the type I errors based on asymptotic and exact approaches can be calculated by finding the tail area identified by the corresponding p-values. \hyperref[fig:surf_g2_m10_n5_A_E]{Figure \ref*{fig:surf_g2_m10_n5_A_E}} to \hyperref[fig:surf_g4_m4_n3_C]{Figure \ref*{fig:surf_g4_m4_n3_C}} and \hyperref[fig:surf_g2_m5_n5_A_E]{Figure \ref*{fig:surf_g2_m5_n5_A_E}} to \hyperref[fig:surf_g4_m3_n2_C_CI]{Figure \ref*{fig:surf_g4_m3_n2_C_CI}} display the surface of type I errors as a function of $\pi$ and $\rho$ when the total sample size is around 30 and 20, respectively. Note that the total number of possible data scenarios increase significantly as the group number increases. Due to the time-consuming nature of computations involved in the E, E+M, and CI approaches, we exclude these methods from consideration and focus solely on the A, C, and M approaches when $g=4$, $m_i=4$, and $n_i=3\; (i=,1,2,3,4)$. According to Tang et al. \citeyearpar{tang2008testing}, a test is classified as liberal if the ratio of the type I error rate to the nominal type I error rate exceeds 1.2. For example, a test is considered liberal if the type I error rate is greater than 6\% given a nominal level of $\alpha=5\%$. On the other hand, a test is considered conservative if the ratio falls below 0.8 or the type I error is less than 4\% given $\alpha=5\%$. Finally, if the ratio is between 0.8 and 1.2, the test is regarded as robust. The asymptotic method produces a large proportion of cases with conservative type I errors that are below the 4\% threshold. The E+M approach performs better than the asymptotic method when the sample size is around 20. However, it performs worse than the asymptotic approach when the sample size is around 30. The C approach has unstable behaviors when the group number ranges from 2 to 4 and becomes extremely conservative when $g=2$. The CI approach produces slightly higher portions of cases with robust type I errors compared to the M method. This is due to the fact that the p-values from the CI approach are determined within the parameter space defined by confidence intervals, which is a subset of the whole parameter space from the M approach. As a result, the p-values obtained from the CI approach are generally smaller than those from the M method, leading to a higher probabilities of rejecting the null hypothesis. In general, although the E approach exhibit a liberal behavior in a small fraction of cases, it controls type I errors within the range of 4\% to 6\% for the majority of scenarios and is considered robust.

\begin{figure}[H]
        \caption{Surface plots of type I errors for $g=2$, $m_i=10$, and $n_i=5$ (M and E+M approaches, $i=1,2$)}
        \label{fig:surf_g2_m10_n5_M_E+M}
     \centering

         \includegraphics[width=\textwidth]{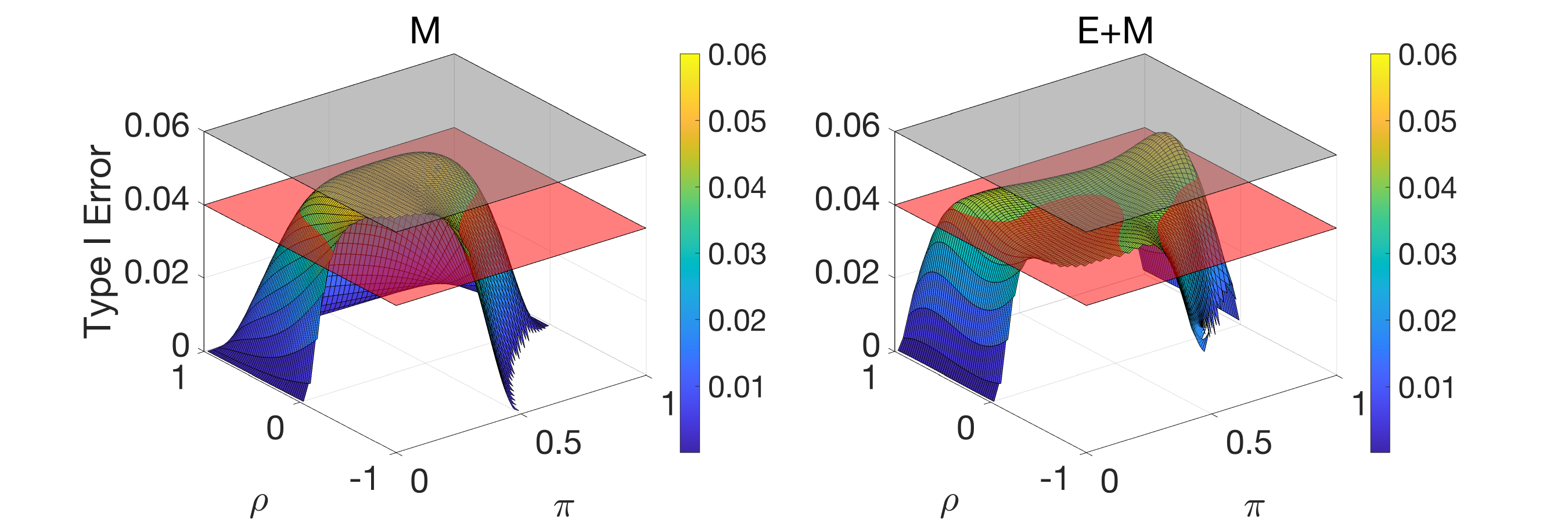}

\end{figure}

\begin{figure}[H]
        \caption{Surface plots of type I errors for $g=2$, $m_i=10$, and $n_i=5$ (C and CI approaches, $i=1,2$)}
        \label{fig:surf_g2_m10_n5_C_CI}
     \centering

         \includegraphics[width=\textwidth]{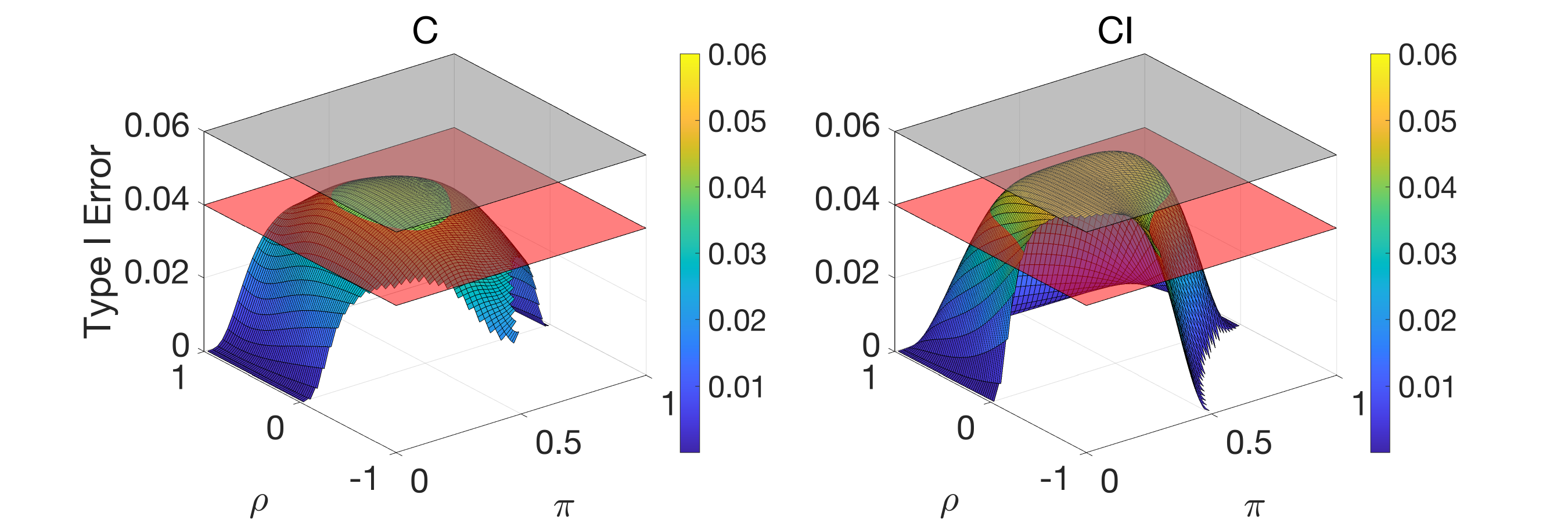}

\end{figure}

\begin{figure}[H]
        \caption{Surface plots of type I errors for $g=3$, $m_i=5$, and $n_i=4$ (A and E approaches, $i=1,2,3$)}
        \label{fig:surf_g3_m5_n4_A_E}
     \centering

         \includegraphics[width=\textwidth]{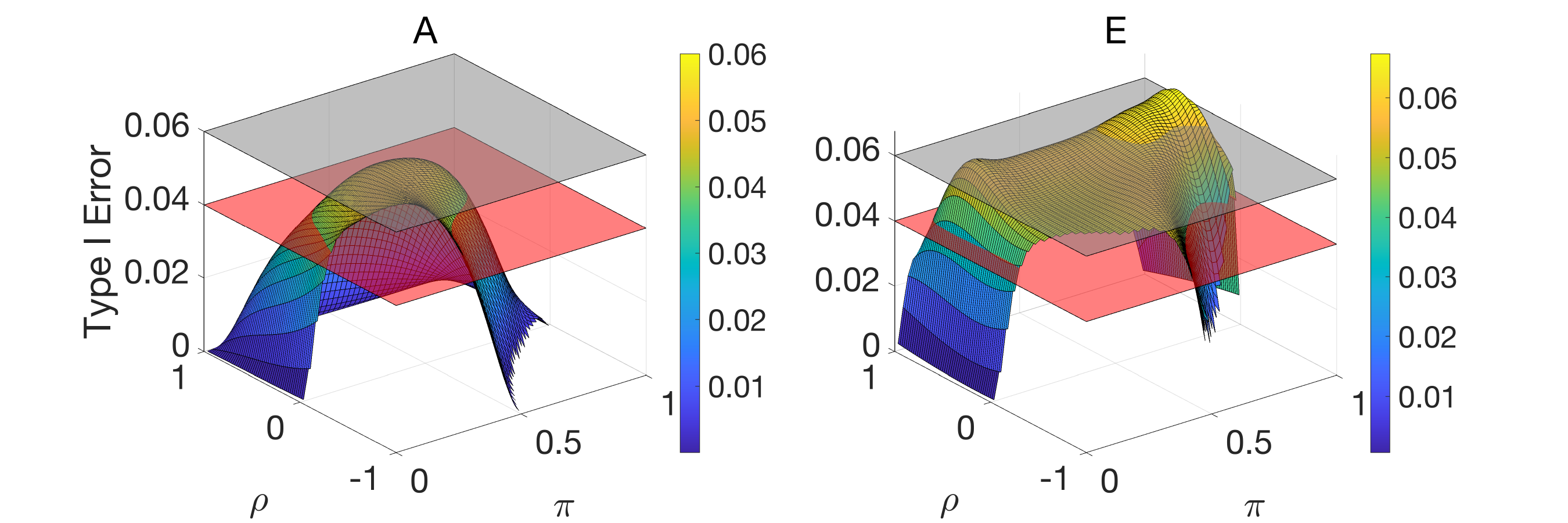}

\end{figure}

\begin{figure}[H]
        \caption{Surface plots of type I errors for $g=3$, $m_i=5$, and $n_i=4$ (M and E+M approaches, $i=1,2,3$)}
        \label{fig:surf_g3_m5_n4_M_E+M}
     \centering

         \includegraphics[width=\textwidth]{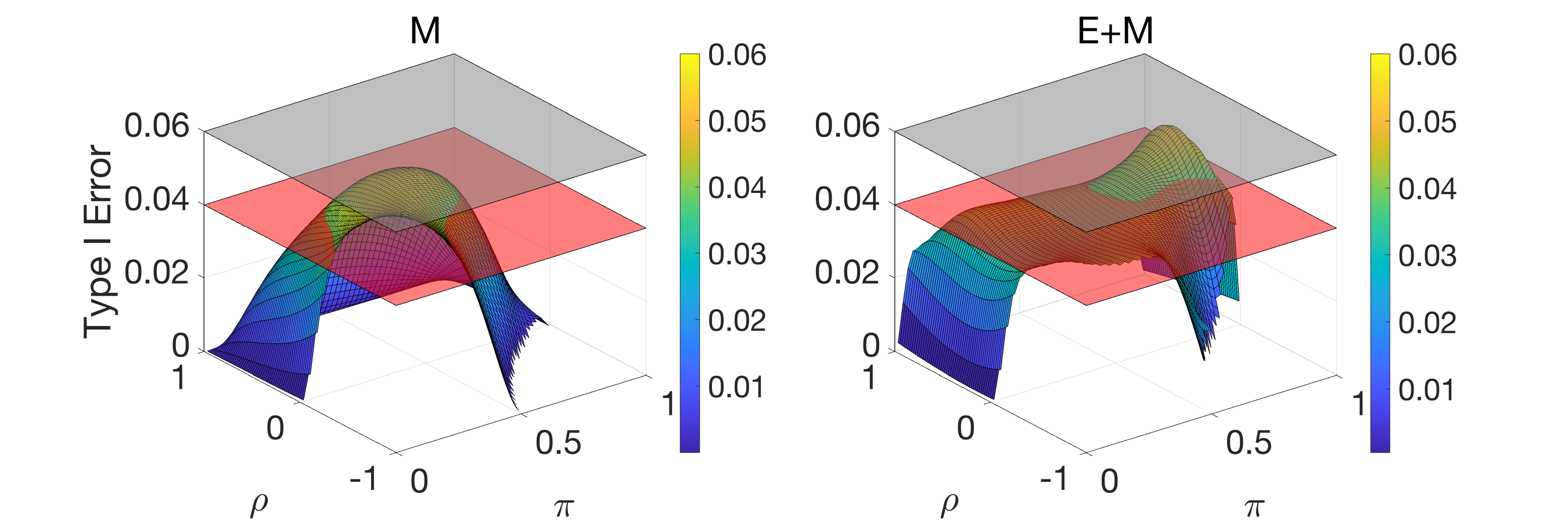}

\end{figure}

\begin{figure}[H]
        \caption{Surface plots of type I errors for $g=3$, $m_i=5$, and $n_i=4$ (C and CI approaches, $i=1,2,3$)}
        \label{fig:surf_g3_m5_n4_C_CI}
     \centering

         \includegraphics[width=\textwidth]{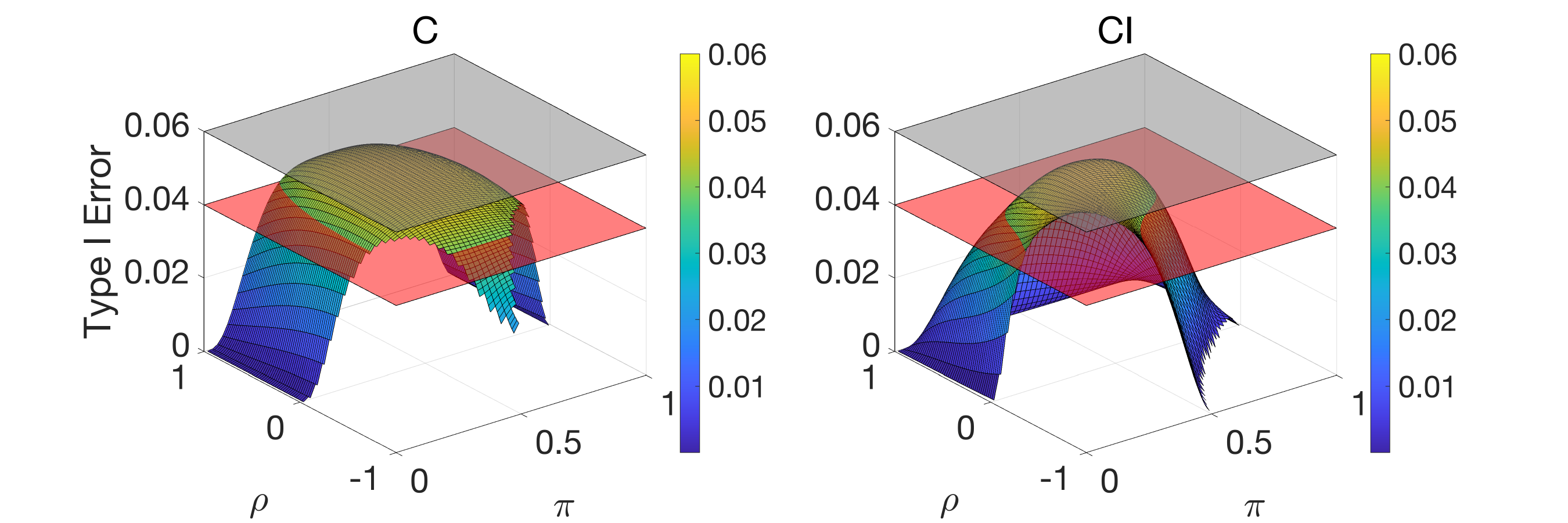}

\end{figure}

\begin{figure}[H]
        \caption{Surface plots of type I errors for $g=4$, $m_i=4$, and $n_i=3$ (A and M approaches, $i=1,2,3,4$)}
        \label{fig:surf_g4_m4_n3_A_M}
     \centering

         \includegraphics[width=\textwidth]{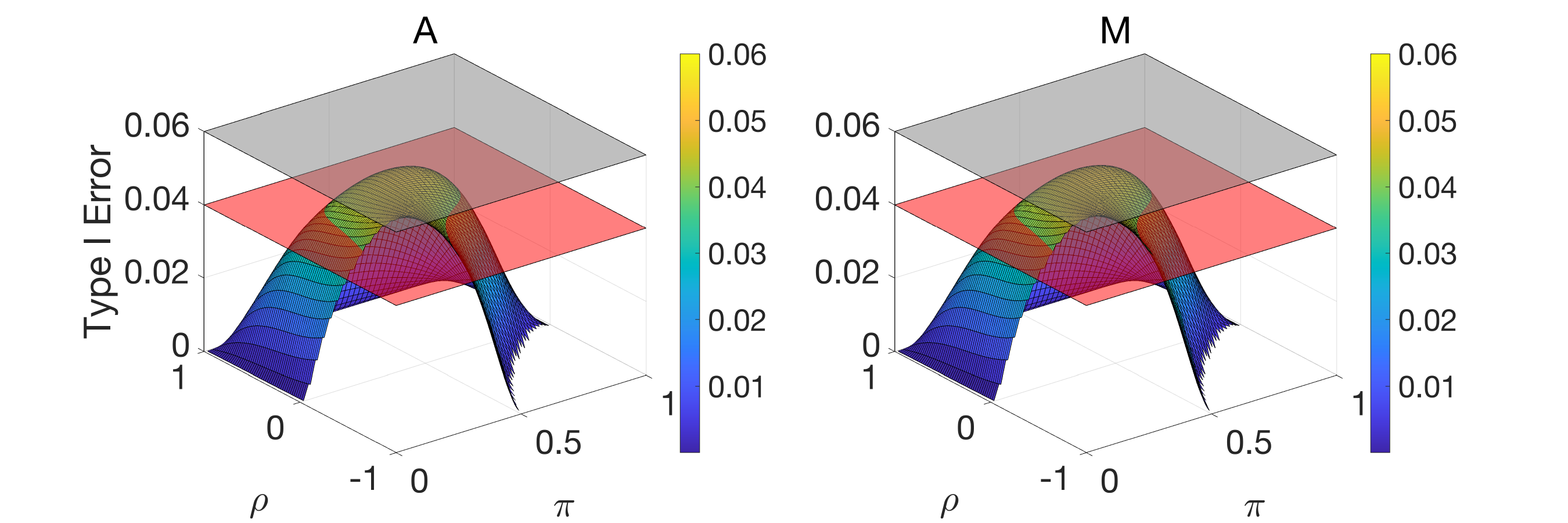}

\end{figure}

\begin{figure}[H]
        \caption{Surface plots of type I errors for $g=2$, $m_i=5$, and $n_i=5$ (M and E+M approaches, $i=1,2$)}
        \label{fig:surf_g2_m5_n5_M_E+M}
     \centering

         \includegraphics[width=\textwidth]{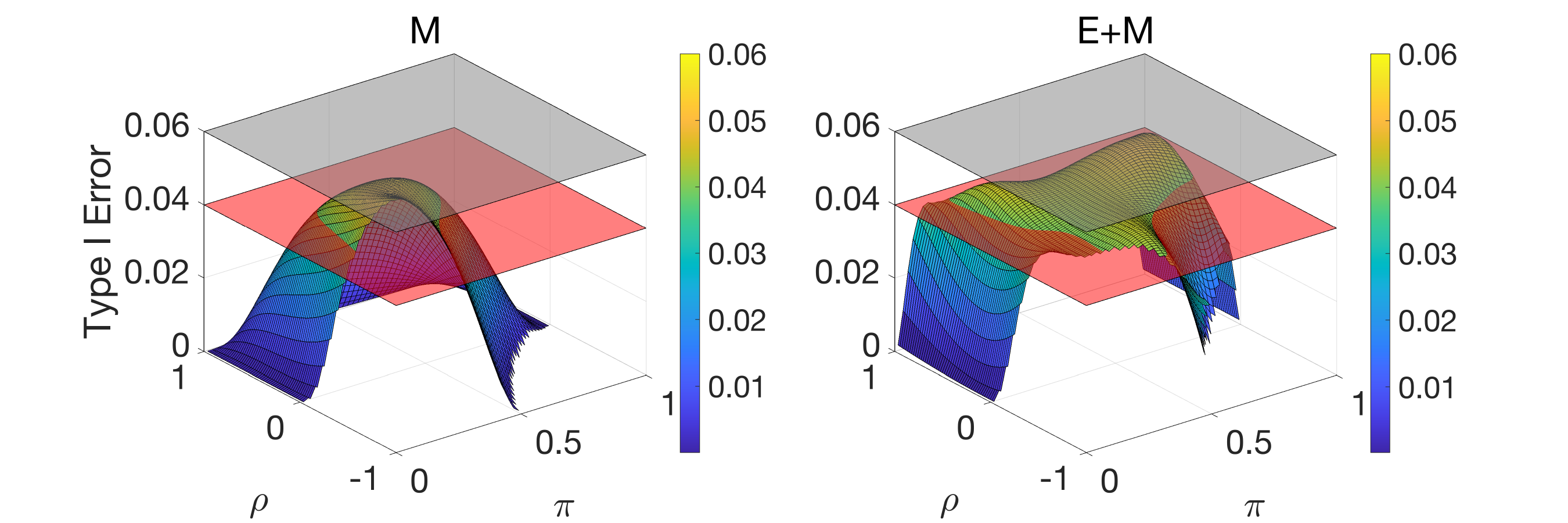}

\end{figure}

\begin{figure}[H]
        \caption{Surface plots of type I errors for $g=2$, $m_i=5$, and $n_i=5$ (C and CI approaches, $i=1,2$)}
        \label{fig:surf_g2_m5_n5_C_CI}
     \centering

         \includegraphics[width=\textwidth]{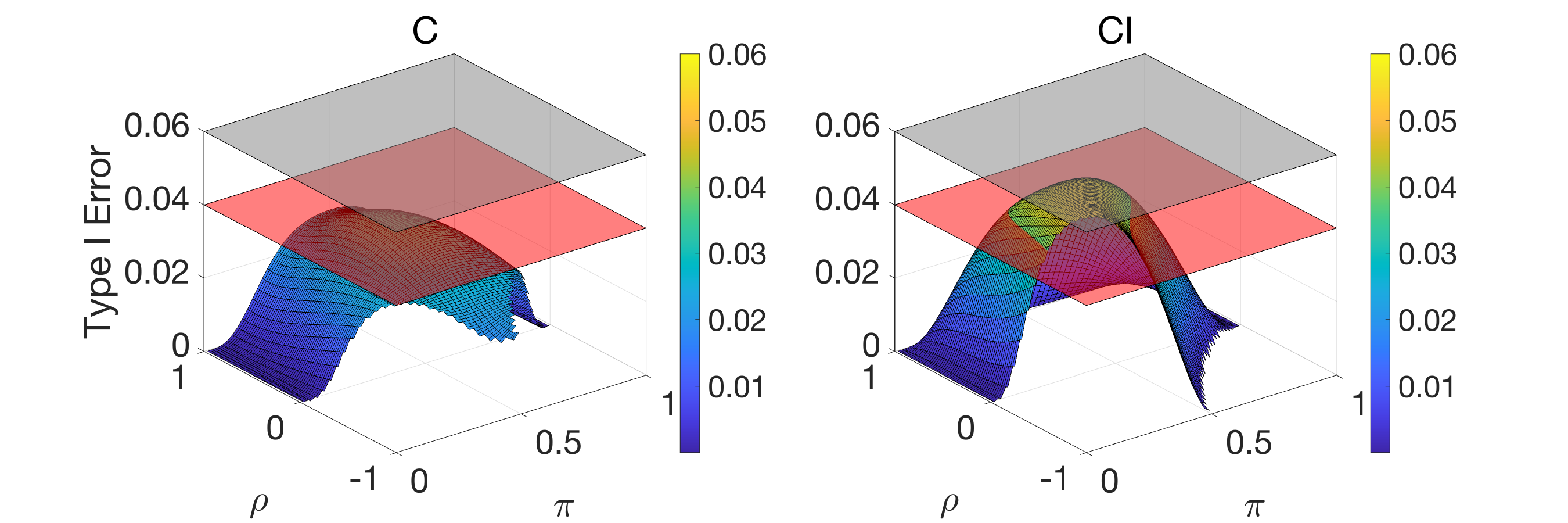}

\end{figure}

\begin{figure}[H]
        \caption{Surface plots of type I errors for $g=3$, $m_i=4$, and $n_i=3$ (A and E approaches, $i=1,2,3$)}
        \label{fig:surf_g3_m4_n3_A_E}
     \centering

         \includegraphics[width=\textwidth]{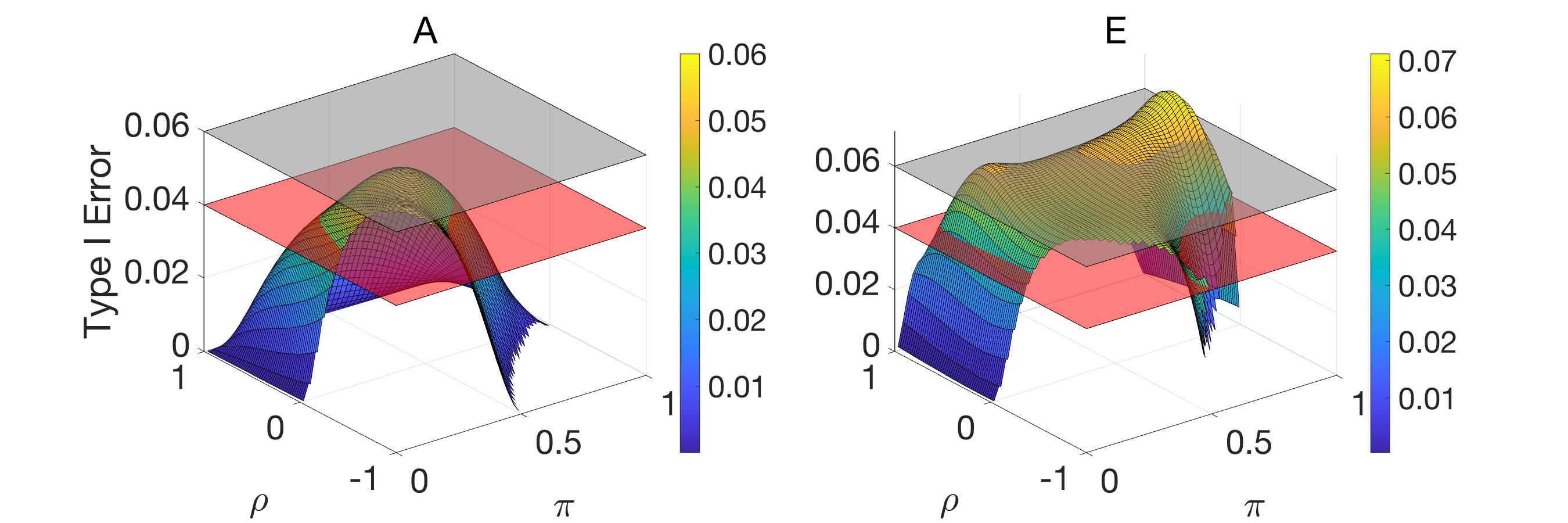}

\end{figure}

\begin{figure}[H]
        \caption{Surface plots of type I errors for $g=3$, $m_i=4$, and $n_i=3$ (M and E+M approaches, $i=1,2,3$)}
        \label{fig:surf_g3_m4_n3_M_E+M}
     \centering

         \includegraphics[width=\textwidth]{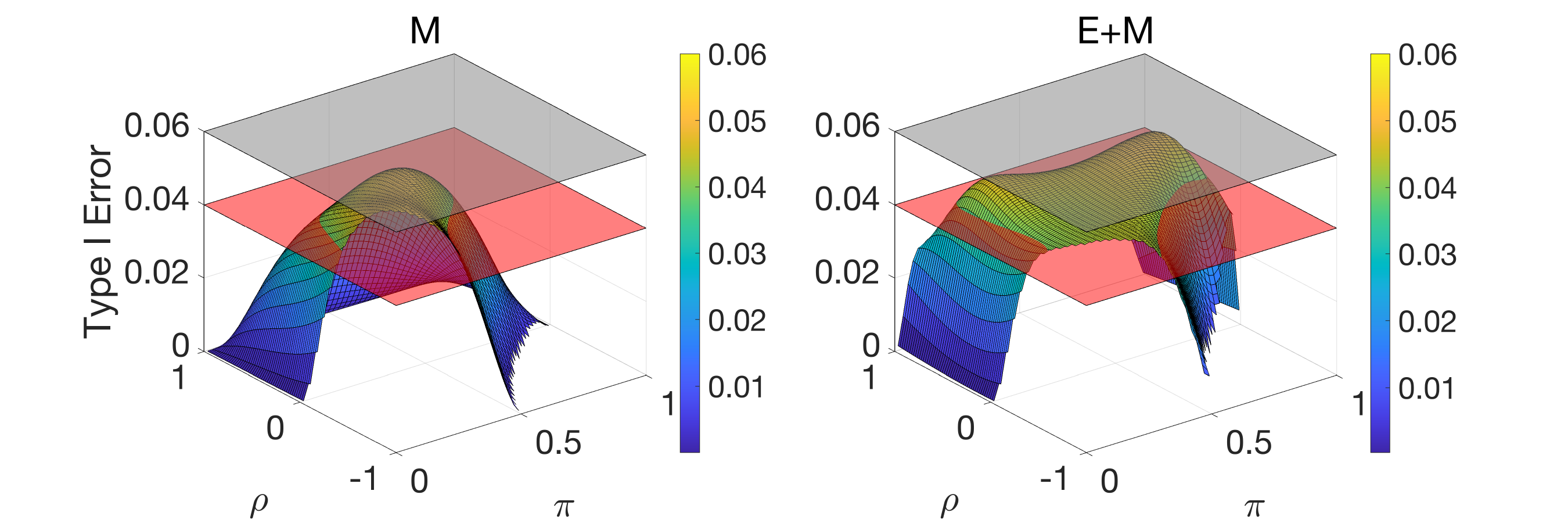}

\end{figure}

\begin{figure}[H]
        \caption{Surface plots of type I errors for $g=3$, $m_i=4$, and $n_i=3$ (C and CI approaches, $i=1,2,3$)}
        \label{fig:surf_g3_m4_n3_C_CI}
     \centering

         \includegraphics[width=\textwidth]{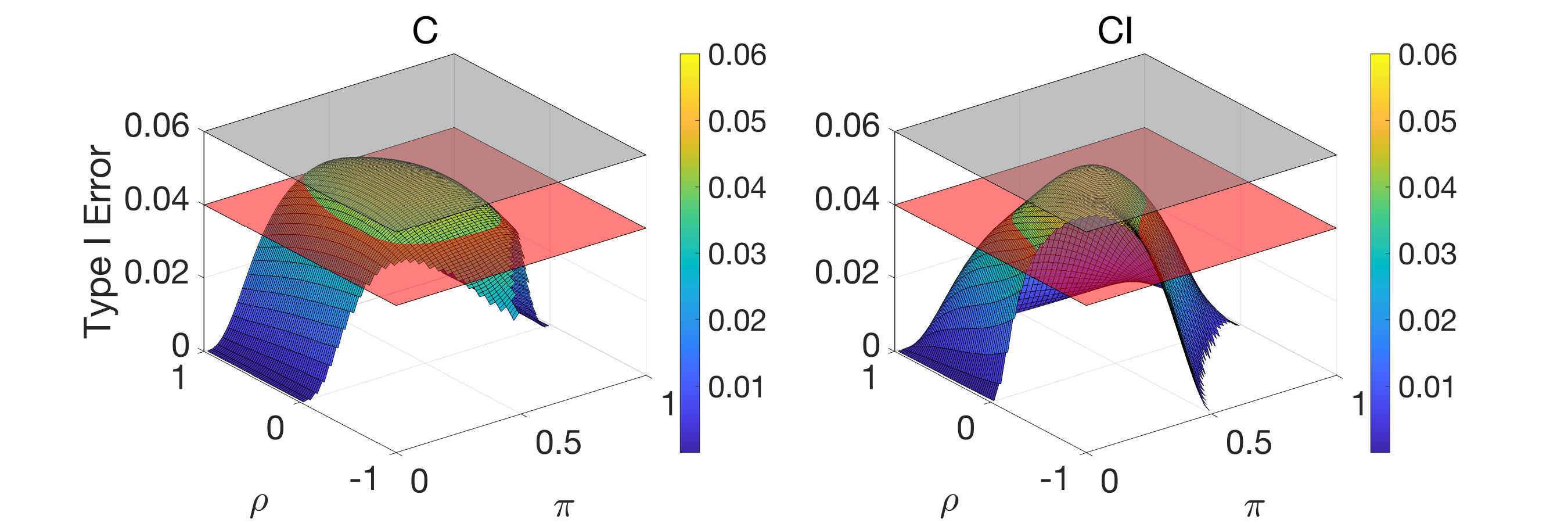}

\end{figure}

\begin{figure}[H]
        \caption{Surface plots of type I errors for $g=4$, $m_i=3$, and $n_i=2$ (A and E approaches, $i=1,2,3,4$)}
        \label{fig:surf_g4_m3_n2_A_E}
     \centering

         \includegraphics[width=\textwidth]{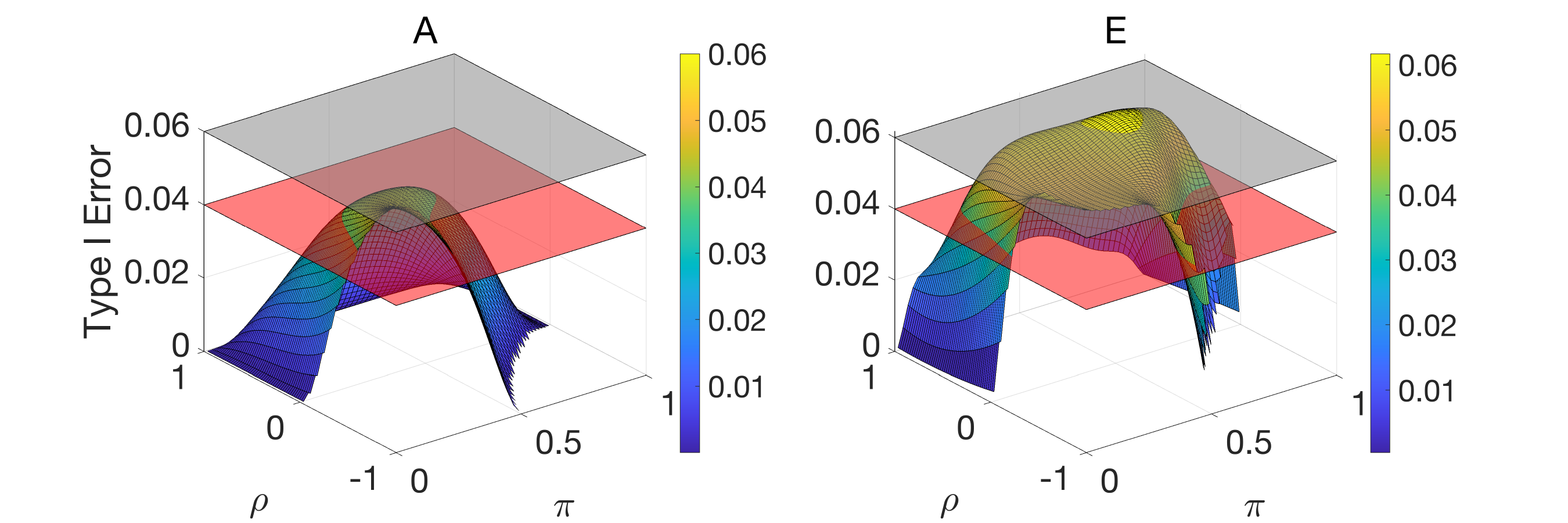}

\end{figure}

\begin{figure}[H]
        \caption{Surface plots of type I errors for $g=4$, $m_i=3$, and $n_i=2$ (M and E+M approaches, $i=1,2,3,4$)}
        \label{fig:surf_g4_m3_n2_M_E+M}
     \centering

         \includegraphics[width=\textwidth]{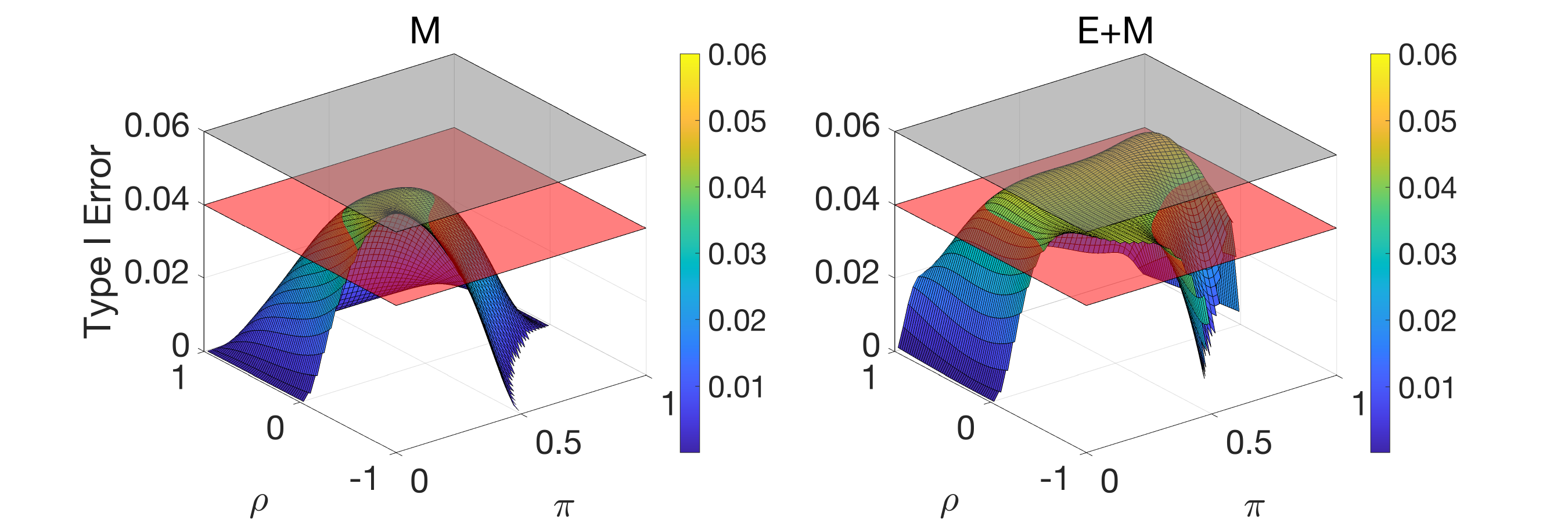}

\end{figure}

\begin{figure}[H]
        \caption{Power plots for $M+N \approx 20$, $\pi_1=...=\pi_{g-1}=0.25$, and $R=1$}
        \label{fig:power_R_10_pi025_20}
     \centering

         \includegraphics[width=\textwidth]{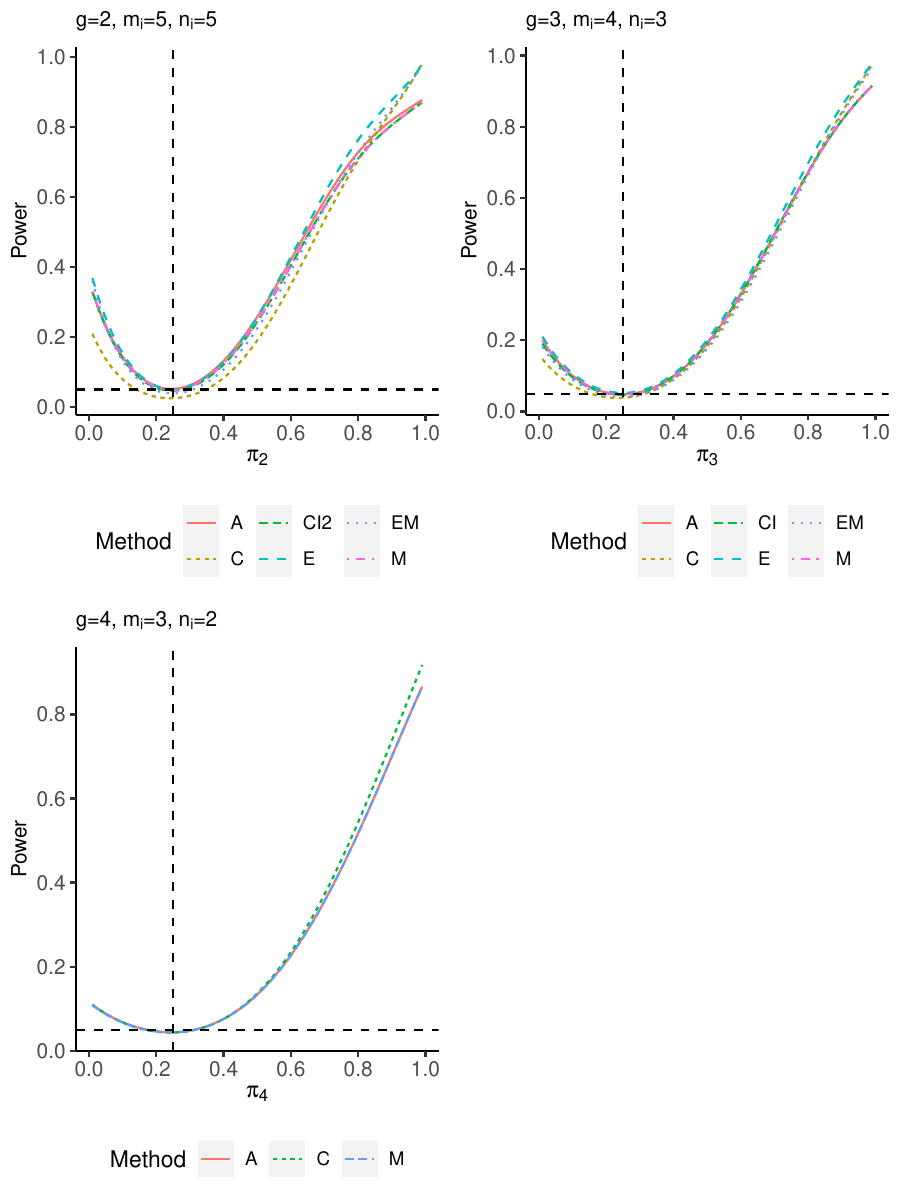}

\end{figure}

\begin{figure}[H]
        \caption{Power plots for $M+N \approx 20$, $\pi_1=...=\pi_{g-1}=0.4$, and $R=1.5$}
        \label{fig:power_R_15_pi040_20}
     \centering

         \includegraphics[width=\textwidth]{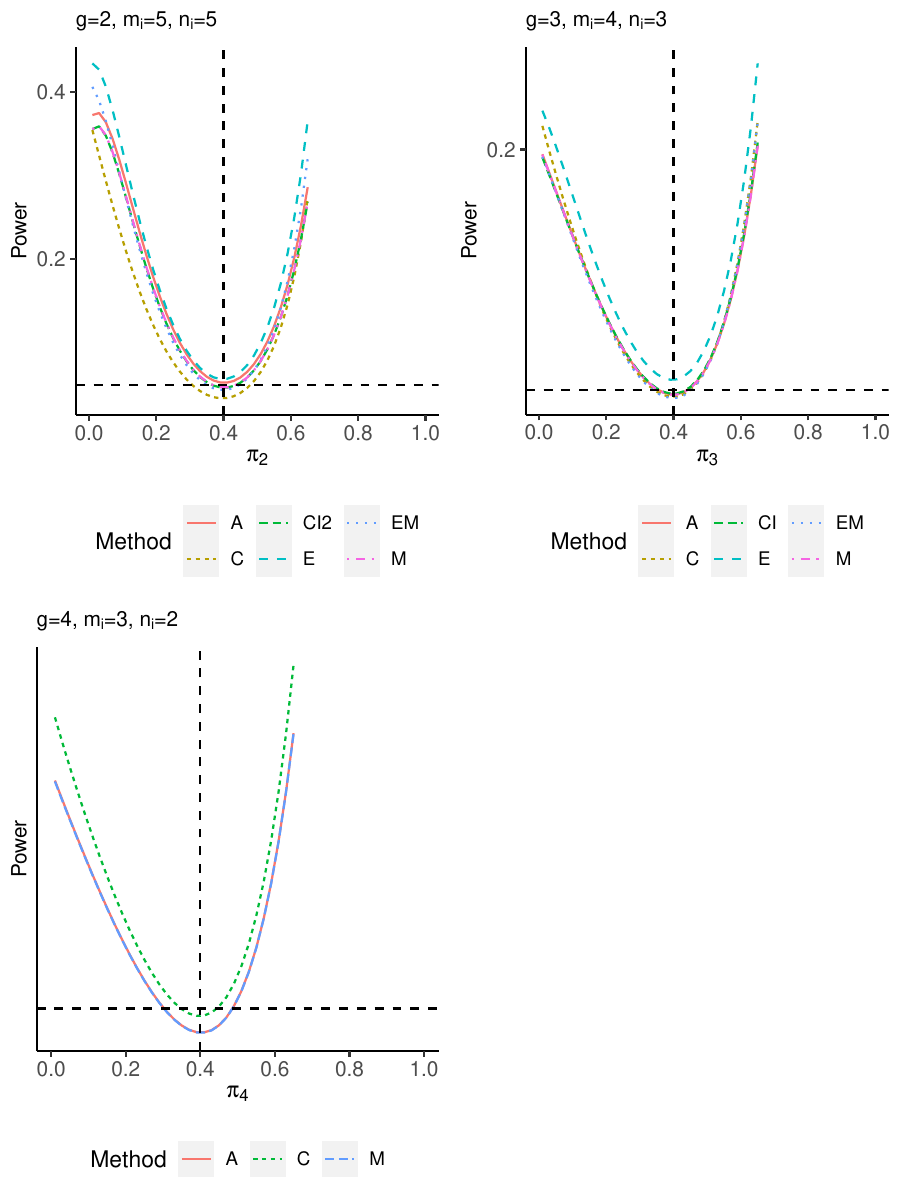}

\end{figure}

\begin{figure}[H]
        \caption{Power plots for $M+N \approx 30$, $\pi_1=...=\pi_{g-1}=0.25$, and $R=1$}
        \label{fig:power_R_10_pi025_30}
     \centering

         \includegraphics[width=\textwidth]{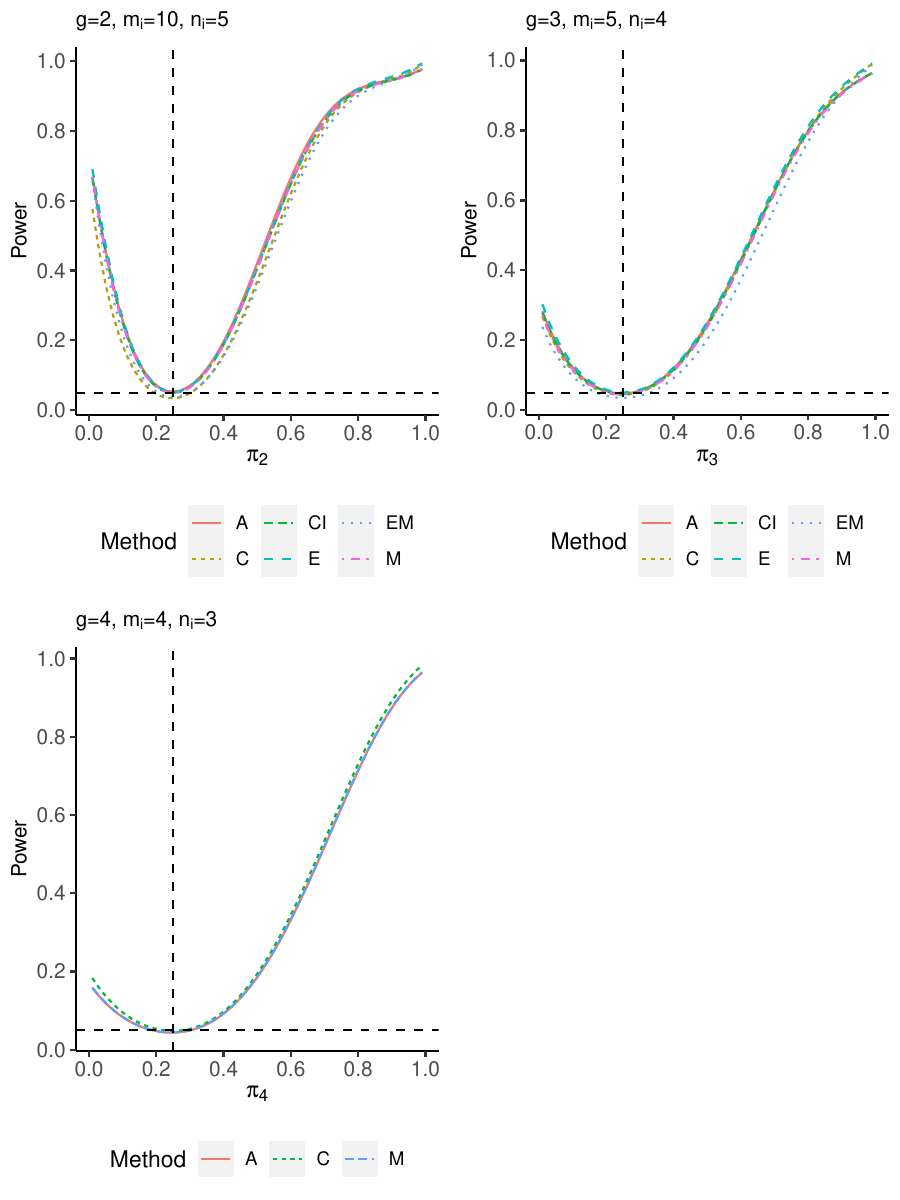}

\end{figure}

\begin{figure}[H]
        \caption{Power plots for $M+N \approx 30$, $\pi_1=...=\pi_{g-1}=0.4$, and $R=1.5$}
        \label{fig:power_R_15_pi040_30}
     \centering

         \includegraphics[width=\textwidth]{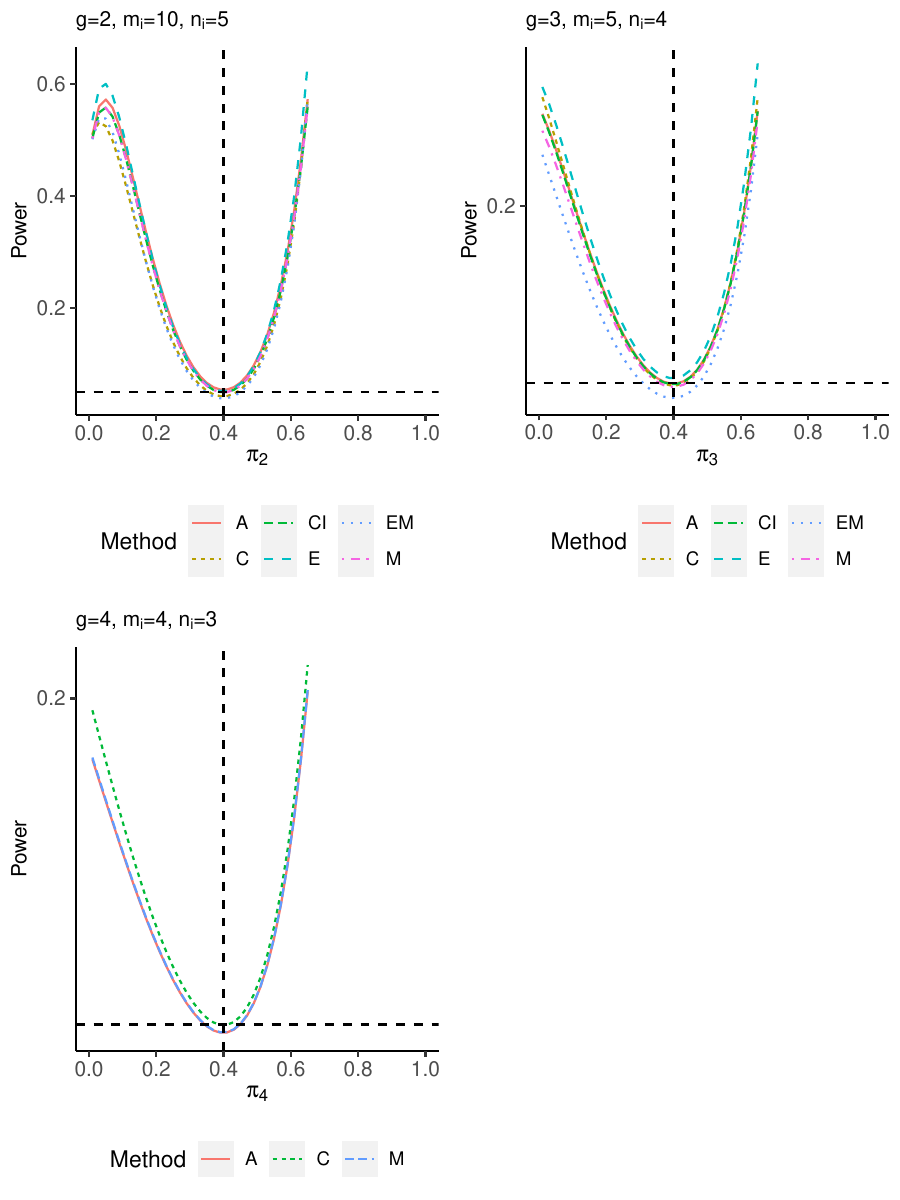}

\end{figure}

Next, we evaluate the power performance with respect to different values of $\pi_g$ while fixing $\pi_1$, $\pi_2$, ..., $\pi_{g-1}$, and $R$. Similar to type I errors, the E, E+M, and CI approaches are excluded when $g=4$, $m_i=4$, and $n_i=3\; (i=,1,2,3,4)$ due to tedious computations. \hyperref[fig:power_R_10_pi025_20]{Figure \ref*{fig:power_R_10_pi025_20}} to \hyperref[fig:power_R_15_pi040_20]{Figure \ref*{fig:power_R_15_pi040_20}} display power plots when the total sample size $M+N$ is approximately 20.  Across all scenarios, the E approach consistently exhibits the highest power. The A, M, and CI approaches yield comparable results regardless of parameter settings. Notably, the C approach demonstrates the lowest power specifically when $g=2$, $m=5$, and $n=5$. As the total sample size approaches 30, all the methods produce increasingly similar powers as indicated in \hyperref[fig:power_R_10_pi025_30]{Figure \ref*{fig:power_R_10_pi025_30}} to \hyperref[fig:power_R_15_pi040_30]{Figure \ref*{fig:power_R_15_pi040_30}}. However, it is worth noting that the E approach continues to maintain the highest power among all methods. The A and M methods also give similar powers as observed when $M+N$ is around 20. For cases where $\pi_1=\pi_2=0.4$, $R=1.5$, $g=3$, $m_i=5$, and $n_i=4$, the CI approach performs slightly better than the M approach. This outcome is expected since $P_{CI}(M^*) \leq P_M(M^*)$, indicating the CI approach is more likely to reject the null hypothesis than the M approach. Overall, all methods demonstrate an increase in power as the total sample size is raised from 20 to 30.

\begin{figure}[H]
        \caption{Power plots for $M+N \approx 20$, $\pi_1=...=\pi_{g-1}=0.25$, and $R=1.5$}
        \label{fig:power_R_15_pi025_20}
     \centering

         \includegraphics[width=\textwidth]{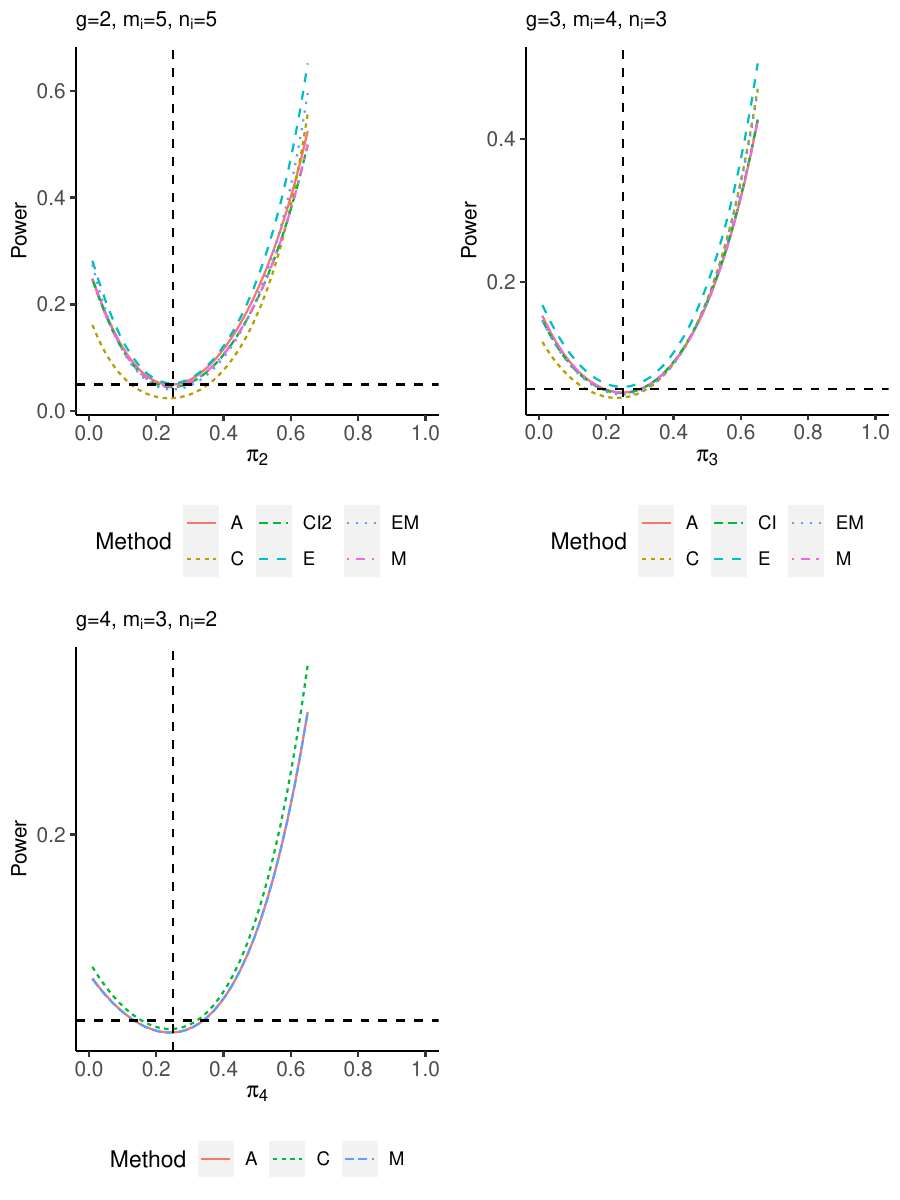}

\end{figure}

\begin{figure}[H]
        \caption{Power plots for $M+N \approx 20$, $\pi_1=...=\pi_{g-1}=0.4$, and $R=1$}
        \label{fig:power_R_10_pi040_20}
     \centering

         \includegraphics[width=\textwidth]{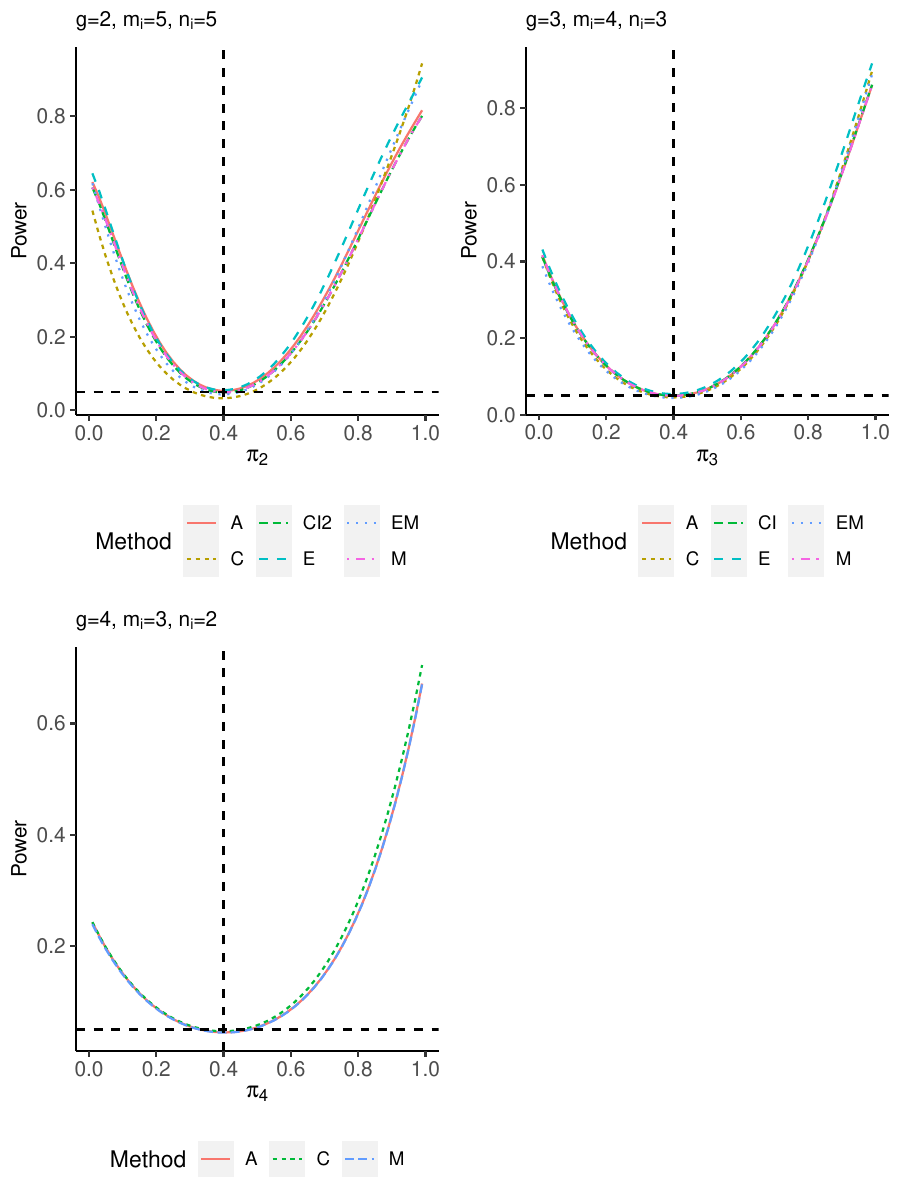}

\end{figure}

\begin{figure}[H]
        \caption{Power plots for $M+N \approx 30$, $\pi_1=...=\pi_{g-1}=0.25$, and $R=1.5$}
        \label{fig:power_R_15_pi025_30}
     \centering

         \includegraphics[width=\textwidth]{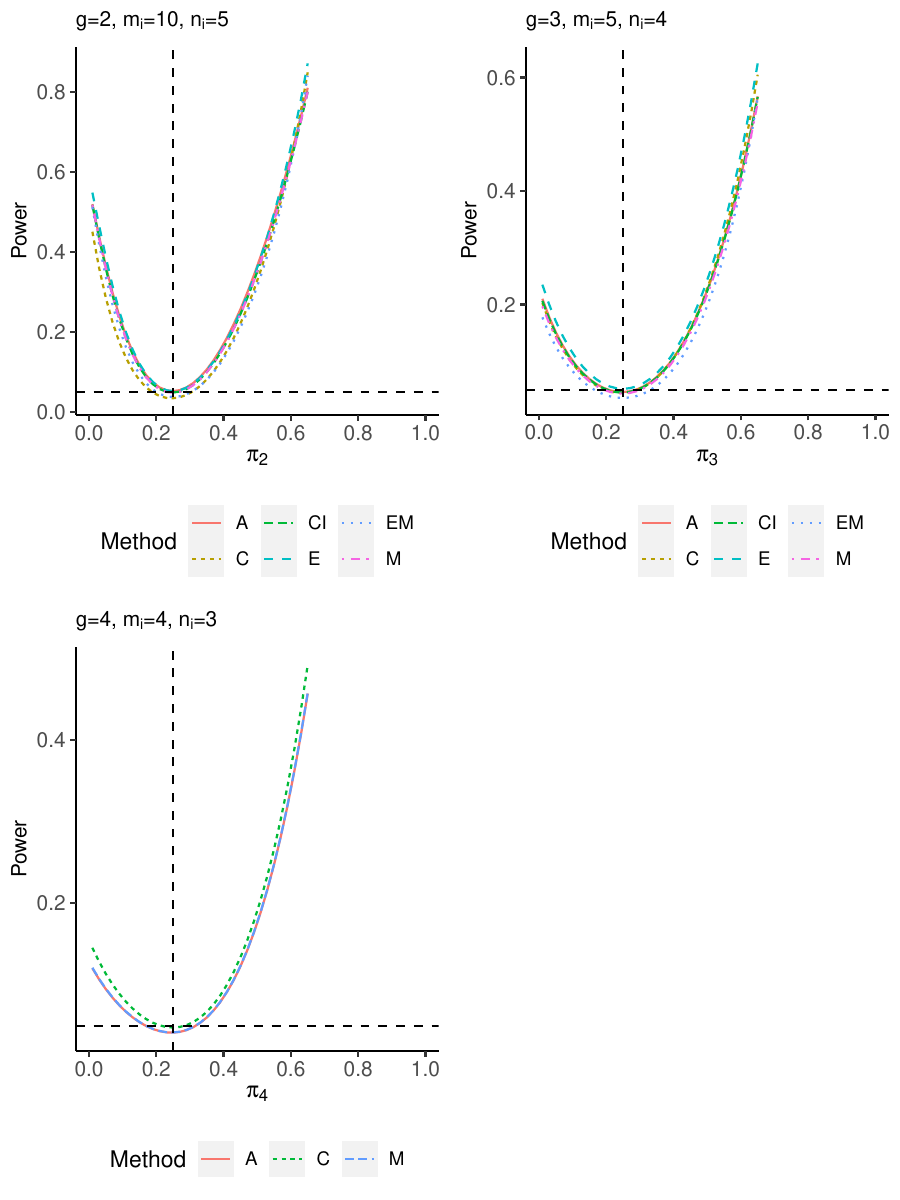}

\end{figure}

\begin{figure}[H]
        \caption{Power plots for $M+N \approx 30$, $\pi_1=...=\pi_{g-1}=0.4$, and $R=1$}
        \label{fig:power_R_10_pi040_30}
     \centering

         \includegraphics[width=\textwidth]{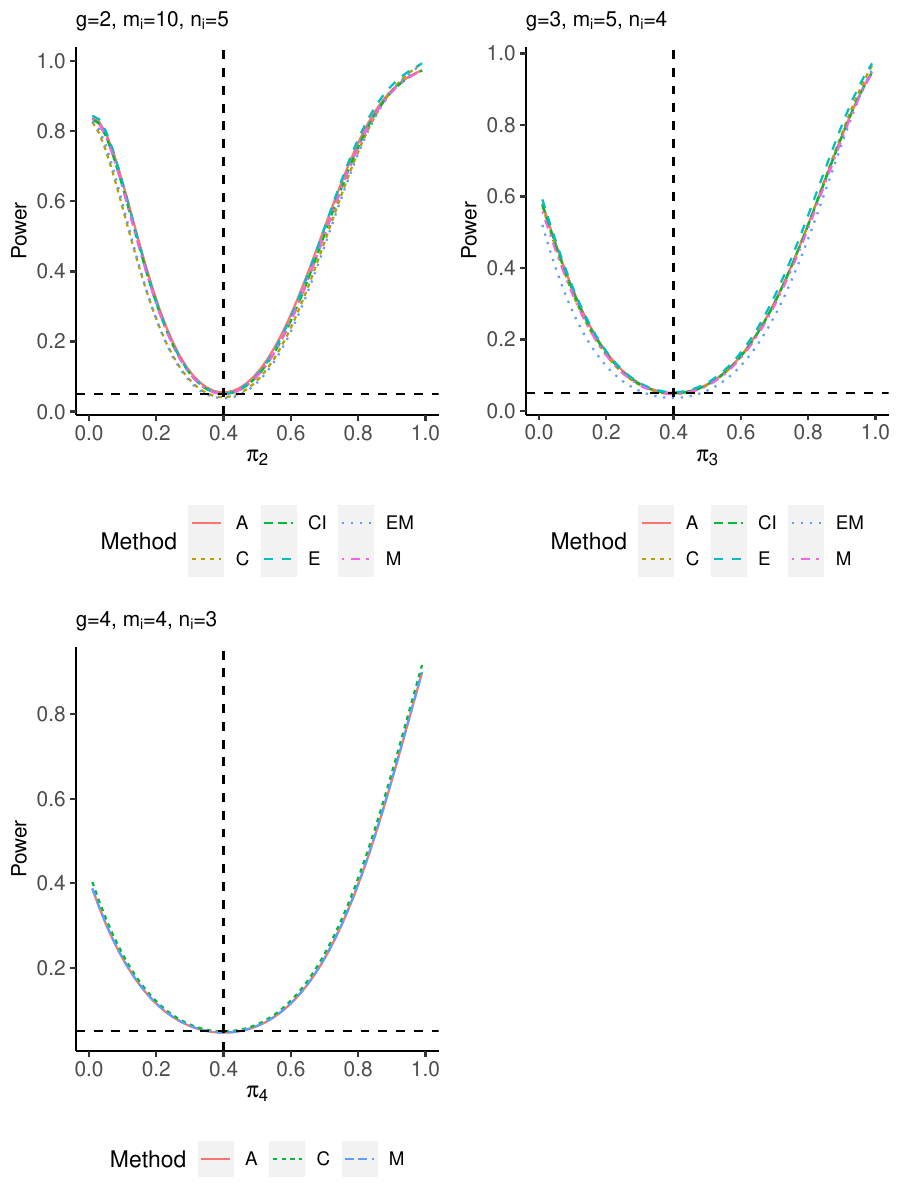}

\end{figure}

\section{Real Examples}
This section provides two practical illustrations of the proposed methods using two real-life examples. The first example pertains to a double-blind randomized clinical trial conducted to investigate acute otitis media with effusion (OME). A total of 214 children with 293 ears who underwent unilateral or bilateral tympanocentesis were randomly assigned to one of the two treatments, cefaclor or amoxicillin (Mandel et al., \citeyear{mandel1982duration}). After the 14-day period, there were 203 evaluable children who met the criteria of not undergoing repeat tympanocentesis, treatment change, or experiencing tympanic membrane perforations. \hyperref[tab:example1]{Table \ref*{tab:example1}} displays the number of children with an age of six years or older at study entry by number of cured ears, treatment received, and disease status at entry (bilateral or unilateral). The main objective is to investigate whether the cure rates of the two treatments are equivalent in this subgroup of subjects with age $\geq$ 6 years. \hyperref[tab:pvalue1]{Table \ref*{tab:pvalue1}} presents the p-values obtained from the asymptotic method and all proposed methods. It is worth noting that all p-values are greater than 0.05, indicating that there is insufficient evidence to reject the null hypothesis at this significance level. 

\begin{singlespace}
\begin{table}[H]
\centering
\caption{The number of children with age $\geq$ 6 years at 14 days by number of cured ears, treatment received, and disease status at entry (bilateral or unilateral)}
\label{tab:example1}
\begin{tabular}{c *{3}{|w{c}{2cm}}  *{2}{|w{c}{2cm}}  }
\hline
\multirow{3}{*}{Treatment} & \multicolumn{3}{c|}{Bilateral at entry}                        & \multicolumn{2}{c}{Unilateral at entry}          \\ \cline{2-6} 
& \multicolumn{3}{c|}{No. of cured ears}& \multicolumn{2}{c}{No. of cured ears} \\ \cline{2-6}  
& 0  & 1 & 2  & 0      & 1       \\ \hline
 Cefaclor & 0 & 1 & 3  & 8       & 11       \\ 
 Amoxicillin & 1  & 0 & 6  & 7     & 11       \\ \hline
\end{tabular}
\end{table}
\end{singlespace}

\begin{singlespace}

\begin{table}[H]
\centering
\caption{P-values based on different approaches for the example in \hyperref[tab:example1]{Table \ref*{tab:example1}}}
\label{tab:pvalue1}
\begin{tabular}{w{l}{3cm}|w{c}{3cm}}
\hline
Approach & p-value \\ \hline
Asymptotic & 0.2257 \\
E & 0.1821 \\
M & 0.2386 \\
E+M & 0.3076 \\
C & 0.3010 \\
CI & 0.2342\\ \hline
\end{tabular}
\end{table}
\end{singlespace}

The second example relates to a retrospective review studying the associations between ocular features and surgical outcomes in infants with stages 4 and 5 retinopathy of prematurity (ROP) (Hartnett \citeyear{hartnett2003features}). Among a total of 35 eyes (22 infants) that were included and underwent surgical procedures, 14 eyes were diagnosed with the zone 1 disease, 20 eyes were diagnosed with zone 2 disease, and the zone of ROP of one eye was unknown due to insufficient information in one patient. The retinal reattachment (Yes/No) was recorded after the first surgery and at the end of follow-up. One of the interests is whether the rates of retinal reattachment are the same between zone 1 and zone 2. According to Hartnett \citeyearpar{hartnett2003features}, 4 eyes (29\%) of the 14 eyes with zone 1 ROP had retinal reattachment, and 13 eyes (65\%) of the 20 eyes with zone 2 ROP had retinal reattachment at the end of follow-up. The two percentages seem to be quite different in magnitude. However, the calculation was based on observations at eye level, which did not account for the between-eye correlation in each individual. \hyperref[tab:example2]{Table \ref*{tab:example2}} displays the distribution of patients by laterality of the disease, zone of ROP, and status of retinal reattachment at the end of follow-up. P-values of various methods are presented in \hyperref[tab:pvalue2]{Table \ref*{tab:pvalue2}} and all the values are greater than 0.05, indicating a failure to reject the null hypothesis that there is no difference in the rate of retinal reattachment between zone 1 and zone 2.

\begin{singlespace}
\begin{table}[H]
\centering
\caption{The number of infants by laterality of the disease, the zone of ROP, and the status of retinal reattachment at the end of follow-up}
\label{tab:example2}
\begin{tabular}{c *{3}{|w{c}{2cm}}  *{2}{|w{c}{2cm}}  }
\hline
\multirow{3}{*}{Zone} & \multicolumn{3}{c|}{Bilateral ROP} & \multicolumn{2}{c}{Unilateral ROP} \\ \cline{2-6} 
 & \multicolumn{3}{c|}{No. of eyes with retinal reattachment} & \multicolumn{2}{c}{No. of eyes with retinal reattachment} \\ \cline{2-6}
 & 0 & 1 & 2 & 0 & 1 \\ \hline
1 & 4 & 1 & 1 & 1 & 1 \\
2 & 1 & 2 & 4 & 3 & 3 \\ \hline
\end{tabular}
\end{table}
\end{singlespace}

\begin{singlespace}

\begin{table}[H]
\centering
\caption{P-values based on different approaches for the example in \hyperref[tab:example2]{Table \ref*{tab:example2}}}
\label{tab:pvalue2}
\begin{tabular}{w{l}{3cm}|w{c}{3cm}}
\hline
Approach & p-value \\ \hline
Asymptotic & 0.4144 \\
E & 0.4513 \\
M & 0.4874 \\
E+M & 0.6310 \\
C & 0.3846 \\
CI & 0.4511\\ \hline
\end{tabular}
\end{table}
\end{singlespace}

\section{Conclusions}
This article focuses on the homogeneity test of prevalence for bilateral and unilateral correlated data with multiple groups. To address the issue of poor type I error control that arises with the asymptotic method when dealing with small sample sizes, we propose five exact tests: E, M, E+M, CI, and C approaches. Exact type I errors and powers are investigated under various parameter settings and different sample sizes.

Numerical studies indicate that the E approach consistently exhibits the strongest control of type I errors. This holds true across different parameter values and sample sizes, outperforming the asymptotic method and other proposed methods. The M and CI approaches generally yield comparable results. The E+M approach performs better than the asymptotic procedure when the sample size is approximately 20. Nevertheless, this superior flips over when the sample size increases to 30. The C approach is unstable across different cases and becomes extremely conservative when $g=2$, $m_i=5$, and $n_i=5$. Regarding statistical powers, the E approach has the highest powers when compared with other methods under all specified settings. Therefore, the E approach is recommended as a result of satisfactory type I error controls and statistical powers. While the exact procedures showcase desirable features, their computation complexity increases very easily if $m_i$ and $n_i$ are not too small. For instance, the total number of possible cases reaches a staggering 12,960,000 when $g=4$, $m_i=4$, and $n_i=3$. Especially for the E, E+M, and CI methods, calculations of p-values are time-consuming as they involve either the computation of $P_E(M^*)$ or the iterative procedures to obtain confidence intervals of $\pi$ and $R$. Addressing this computational challenge is an area that remains open for future research.

\bibliography{ref}

\end{document}